\documentclass[12pt]{article}
\usepackage{fullpage,epsfig,graphics,amsbsy,amssymb,cancel}
\usepackage{graphicx}
\usepackage{amsfonts}
\usepackage{amssymb,amsmath,amscd}
\usepackage{feyn}

\newcommand{\BS}{\boldsymbol}
\newcommand{\BB}{\mathbb}

\newcommand{\bea}{\begin{eqnarray}}
\newcommand{\eea}{\end{eqnarray}}
\newcommand{\beq}{\begin{eqnarray}}
\newcommand{\eeq}{\end{eqnarray}}
\newcommand{\nn}{\nonumber}
\newcommand{\Tr}{\textrm{Tr}}

\begin{document}
\setcounter{page}{0}
\thispagestyle{empty}
\begin{flushright} \small
UUITP-23/09  \\
 \end{flushright}
\smallskip
\begin{center} \LARGE
{\bf Odd  Chern-Simons Theory,  Lie Algebra Cohomology and Characteristic Classes}
 \\[12mm] \normalsize
{\large \bf Jian~Qiu and Maxim~Zabzine} \\[8mm]
 {
 \it
   Department of Physics and Astronomy,
     Uppsala university,\\
     Box 516,
     SE-75120 Uppsala,
     Sweden\\}
\end{center}
\vspace{10mm}

\begin{abstract}
 \noindent
  We investigate the generic 3D topological field theory within AKSZ-BV framework.
We use the Batalin-Vilkovisky (BV) formalism to construct
explicitly cocycles of the Lie algebra of formal Hamiltonian
vector fields and we argue that the perturbative partition function gives rise to secondary characteristic classes.
 We investigate a toy model which is an odd analogue
of Chern-Simons theory, and we give some explicit computation of
two point functions and show that its perturbation theory is
identical to the Chern-Simons theory. We give concrete example of
the homomorphism taking Lie algebra cocycles to $Q$-characteristic
classes, and we reinterpreted the Rozansky-Witten model in this
light.

\end{abstract}


\eject \normalsize \eject

\tableofcontents

\eject

\section{Introduction}
\label{intro}

Topological field theory (TFT) a is well-developed subject
spreading across physics and
 mathematics. TFT can be viewed as a very powerful machine for producing the topological
  invariants. If one looks at TFT from the point of view of path integral, then one should
   deal with the appropriate gauge symmetries and thus with BRST formalism.
 A usual way of constructing a topological
field theory is that one proposes a set of BRST transformations
for a set of fields, and then write down an action which is
usually a BRST-exact term plus perhaps some additions of
topological nature (e.g., such as the pull back of the K\"ahler form of
the target manifold). Apart from the insight required to come up
with a reasonable BRST rule, one is constantly faced with the
problem that the BRST transformation closes only on-shell, and the
problem of determination of observables etc. Thus dealing with all these issues
 is somewhat ad hoc.

The Alexandrov-Kontsevich-Schwarz-Zaboronsky (AKSZ) construction \cite{Alexandrov:1995kv}
  is an elegant and powerful tool to  engineer the topological field theories in
various dimensions within the BatalinÐVilkovisky (BV) formalism.
Many problem are all
avoided with one single ingenious stroke of the AKSZ construction.
Its beauty lies in that it converts the finding of the BRST
transformation rules to a purely geometrical problem, namely, one
seeks the so called $Q$-structure on a target manifold. The
$Q$-structure is by definition an odd nilpotent vector field. This
does not seem much improvement so far, but with the unifying
language of graded manifolds (GrMfld), the possible $Q$-structures
are well understood. For example, on a degree 1 GrMfld, a
$Q$-structure encodes the data of a Lie algebroid. Thus the BRST
rule will be related to the Lie algebroid differential for the
target manifold,  e.g. see \cite{future} for the construction of a
whole gamut of topological models. The AKSZ construction is done
naturally within the BV formalism, which then clarifies the
problem of on-shell closure of BRST transformation and at the same
time gives geometrical interpretation to the otherwise
unilluminating routine of gauge fixing.

On the other hand, in physics we are equipped with the handy tool
of path integral which, albeit being totally formal, allows one to
manipulate the formalisms conveniently. And it is no new
phenomenon that one could use a topological field theory and path
integral to produce non-trivial mathematical results.  In this work we offer the systematic
 study of the perturbative AKSZ-BV topological theories. Moreover we suggest the interpretation of
  the perturbative correlators and partition function in these theories. In particular we concentrate
  our attention on 3-dimensional (3D) theories.

The present
work is heavily influenced by several pieces of work along this
direction. First the Chern-Simons perturbation theory
\cite{Axelrod:1991vq}, where the evaluation of the partition
function led to the physical construction of invariants of
3-manifolds. Later Kontsevich \cite{Kontsevich:Feynamn} exposed the connection between the
Feynman integral and graph (co)homology (namely the Feynman
integral gives a cocycle in the graph complex); and thereby the
construction of the low dimension topological invariants. Another
piece of inspiration came from the works of Schwarz \cite{Schwarz:1999vn} and
  Lazarev and Hamilton \cite{Hamilton:2007tg}, especially the latter, who used the tool of BV path
integral to furnish a proof of the claim made by Kontsevich. Their
proof made an excursion of first showing that the path integral is
a cocycle in the cohomology of the Lie algebra of Hamiltonian
vector fields. Since the latter is proven to be isomorphic to the
graph cohomology, one can first send a graph chain to an element
in the Lie algebra chain complex, then evaluate this chain in the
path integral giving the desired cochain.
  We will show that all these ideas arise naturally within AKSZ-BV framework.
   Indeed BV path integral always give rise to a certain cocyles and the perturbative theory
    offers the concrete way of calculating them.  Although we look mainly at 3D AKSZ models,
     many ideas can be extended to other AKSZ  theories.

Being furnished with a cocycle coming from BV path integral one is led naturally to
  construct some characteristic classes using the Chern-Weil homomorphism.
Now instead of plugging in the curvature two form to an invariant
polynomial of Lie algebra, we plug in a flat connection into a cocycle.
 This is exactly what happens when we calculate the partition function of the  AKSZ theory.

One purpose of this work is clarify what exactly the
perturbation theory of these AKSZ models is computing. The
partition  function for such models turns out to be the
(hopefully non-vanishing) characteristic classes of the relevant
$Q$-(super)manifold. In the work by Lyakhovich, Mosman and Sharapov
\cite{lyakhovich-2009}, they are able to use graph cohomology\footnote{Their graph complex is slightly different
 from what we consider  and it  is isomorphic to
the cohomology of Lie algebra of formal vector fields vanishing at
the origin.} to find three infinite series of characteristic
classes of any $Q$-manifold. Especially, their B,C series depend
on the properties of the homological vector $Q$ alone and survive
even for a flat manifold. In a nut shell, due to the observation
$L_Q\partial_i\partial_j Q^k=0$ where $Q^i\partial_i$ is a
homological vector field, they show if one plugs the second Taylor
coefficient of $Q^i$ into certain graphs made of 3-valent
vertices, out comes some $Q$-characteristic classes.\footnote{This
gives their B,C series of invariants, their A series come from
two valent graphs and requires the vanishing of Pontryagin class.}
This version of the characteristic classes for the flat bundle is again tied to a
second construction of graph cycle (except they are using it
backwards) by Kontsevich. The construction is intuitive, one
obtains a graph cocycle by plugging into the vertices the Taylor
coefficient of the Hamiltonian lift of $Q$ and connecting edges
using the symplectic form. We shall show that this is indeed what
happens when one evaluates the partition function for the AKSZ
models. For such model the interaction term is just the
Hamiltonian lift of $Q$, and for anyone who knows anything about
perturbation theory in physics, the evaluation of the Feynman
diagrams are just about plugging the Taylor coefficients of the
interaction terms.

The article is organized as follows, the BV formalism is reviewed
in section \ref{sec_BV}. We also show that the quantum observables form a
closed algebra and the path integral gives a cocycle in Lie
algebra cohomology of formal Hamiltonian vector fields generated
by these observables. In section \ref{sec_review} we review some relevant background material.
 The characteristic classes of
 flat bundles  are recalled and we discuss the scenario in which
 they can arise in path integral.  The isomorphism
between Lie algebra (co)homology and graph (co)homology is
 sketched without any claim for rigor.
We give the construction of the
AKSZ model in section \ref{sec_A_3D_Model}, in particular, the free
theory gives a cocycle in Lie algebra cohomology of formal
Hamiltonian vector fields of the target space.
To do serious perturbation calculation,  one needs to gauge fix the
model; this is the topic of section \ref{sec_gauge_fix}.  There we
 present the set of Feynman rules and we investigate the perturbative
  partition function. We claim that the partition function corresponds to
 a characteristic class of appropriate flat bundle.    Sections \ref{sec_Flat_Bundle}-\ref{sec_RW_model}
 are dealing with different examples of 3D AKSZ models.
In section \ref{sec_Flat_Bundle} we consider the AKSZ model associated to the $Q$-equivariant
 vector bundle.
In section \ref{sec_Min_model} we examine  3D AKSZ model constructed on a flat symplectic space
$\mathbb{R}^{2m}$, and we show that it is a kind of odd analogue
of Chern-Simons perturbation theory and the weight functions
associated with the diagrams are identical to
Chern-Simons and the Rozansky-Witten model.
Finally as a grand
finale section \ref{sec_RW_model}, we put all the ingredients together
and reformulate the Rozansky-Witten model in the light of Lie
algebra cohomology and the characteristic classes of flat bundles.
 At the end of the paper there are two appendices which contains some
  technical calculations relevant for the paper.

\section{BV Formalism}\label{sec_BV}%

We give the essential facts about BV formalism in this section and
show that the standard manipulations in the BV framework allow us
to interpret the path integral as a certain cocycle.

The original BV formalism was for the supermanifolds, namely
manifolds with $\mathbb{Z}_2$ grading, yet the formalism may be
carried to $\mathbb{Z}$-graded manifolds making some of the
results stronger. A degree $n$ graded manifold is by definition
locally parameterized by coordinates of degrees 0 up to $n$. And
these coordinates are glued together through \emph{degree
preserving} transition functions (for more details on the graded
manifolds, see \cite{Voronov:2001qf} and
\cite{Roytenberg:2002nu}). An example of such a manifold is:
$T[1]M$; the notation being: $M$ is an ordinary manifold, $T[1]$
means that we take the total space of the tangent bundle of $M$
and we assign the fiber coordinate degree 1. This is an odd
manifold since the highest fiber degree is 1. An example of even
graded manifolds is $T^*[2]T[1]M$, locally, we have $x^{\mu}$ as
the coordinate of $M$, the coordinate $v^{\mu}$ parameterizing the
fiber of $T[1]M$ is given degree 1, the coordinates dual to
$x^{\mu}, v^{\mu}$ are $P_{\mu}$ and $q_{\mu}$ of degree
$-\deg(x)+2=2$ and $-\deg(v)+2=1$ respectively. The advantage of
using graded manifolds instead of supermanifolds is that degrees
of these coordinates eventually correspond to the ghost number in
a physical theory.

The BV manifold is a manifold where the space of functions is equipped
 with the structure of BV algebra which is defined as the Gerstenhaber algebra (odd Poisson algebra)
  together with an odd Laplacian.
  Simply speaking the BV manifold is a manifold equipped with an odd symplectic form.
The archetypical example of such spaces is of the form
$T^*[-1]M$, where $M$ itself is allowed to be a graded manifold.
The reason for the degree $-1$ shift is to make the BRST
transformation of ghost number $+1$ in the end. For definiteness,
let us take the coordinate of $M$ as $x$ and that of the fiber
$x^+$, then the canonical symplectic form of the BV space is just
$\omega=dx\wedge dx^+$.

If $M$ has dimension $n$, then a Lagrangian submanifold
(LagSubMfld) ${\cal L}$ is a dimension $n$ submanifold of the BV
space such that $\omega|_{{\cal L}}=0$. Suppose that a volume form
$\mu(x)$ is given for $M$, then we have also a volume form for
$T^*[-1]M$ which is $\mu^2(x)\wedge^{n}dx^+\wedge^{n}dx$. With the
density $\mu(x)$
we can define a Laplacian%
\bea\Delta \equiv \frac{1}{\mu^2(x)}\frac{\partial}{\partial
x}\mu^2(x)\frac{\partial}{\partial x^+}\ ,\nn\eea%
which can be checked to satisfy $\Delta^2=0$.

The key fact of the BV formalism \cite{Schwarz:1992nx} is the statement that the
integral of a function $f$ over a LagSubMfld is invariant under
continuous deformation of the LagSubMfld provided $f$ satisfies
$\Delta f=0$; and the integral of $\Delta$-exact functions gives
zero. This statement is just the Stokes theorem in disguise
\cite{Witten:1990wb}. By Fourier transforming the odd degree
coordinates in $T^*[-1]M$ (namely, exchanging the coordinate and
its dual momentum), the Laplacian $\Delta$ becomes the de Rham
differential $d$ over the degree even submanifold of $T^*[-1]M$.
And the integration of functions over LagSubMfld is reformulated
as integration of forms along submanifolds. In contrast to $d$,
$\Delta$ is not a derivation (does not obey the Leibnitz rule),
in fact, when acting on a product of functions, it gives%
\bea
\Delta(fg)=(\Delta f)g+(-)^{|f|}f(\Delta
g)+(-)^{|f|}\{f,g\}\label{LAP_BV_GR_BR}~,
\eea%
where $\{\cdot,\cdot\}$ is the odd Poisson bracket corresponding
to the odd symplectic form $\omega$.

We are going to explore the consequence of (\ref{LAP_BV_GR_BR}).
The usual use of BV formalism is in the quantization of gauge
system: suppose one has an action $S$ satisfying $\Delta
e^{-S}=0$, then one seeks a suitable ${\cal L}$ such that the
restriction of $S$ to ${\cal L}$ has a non-degenerate quadratic
term. The choice of the LagSubMfld is the choice of the gauge
fixing condition, and due to $\Delta e^{-S}=0$, the end result
should not depend on the choice of gauge fixing. Having chosen
${\cal L}$, one then inserts operators ${\cal O}$ with
$\Delta({\cal O}e^{-S})=0$ into the path integral and obtain the
expectation value of ${\cal O}$. It is usually stated that the
path integral is a homomorphism sending elements of
$H(T^*[-1]M,\Delta_q)$ ($\Delta_q \equiv e^S\Delta e^{-S}= \Delta - \{S, ~\}$) to the
number fields. Due to the fact that $\Delta$ is not a
derivation, there is no ring structure defined for the
cohomology group $H(T^*[-1]M,\Delta_q)$. This point of view is of
course correct, however, it misses some rich structure innate in
the BV formalism. In fact the cohomology group of $\Delta$ is
quite boring, as $\Delta$ can always be Fourier transformed into a
de Rham differential. One of the purposes of this paper is to
elaborate some results in the paper by Schwartz
\cite{Schwarz:1999vn} and by Hamilton and Lazarev
\cite{Hamilton:2007tg}. The first crucial observation made by
Schwartz is that the quantum observables (namely functions
satisfying $\Delta_qf=0$) form a closed algebra under the Poisson
bracket, more concretely, by using  (\ref{LAP_BV_GR_BR})
 \bea
\{f,g\}&=&(-1)^{|f|}\Delta(fg)-(-1)^{|f|}(\Delta f)g-f(\Delta g)\nn\\
 \label{bracketquantumus}&=&(-1)^{|f|}(\Delta_q(fg)+\{S,fg\})-(-1)^{|f|}\{S,f\}g-f\{S,g\}=(-1)^{|f|}\Delta_q(fg)~, \eea%
hence the bracket quantity $\{f,g\}$ remains closed under
$\Delta_q$. But the bracket here does not yield a super Lie
algebra structure for the quantum observables: the difference
between the two is a shift in the assignment of the degree. More
concretely, the Poisson bracket appearing here is odd and obeys
$\{f,g\}=-(-1)^{(|f|+1)(|g|+1)}\{g,f\}$, while for a super Lie
algebra we would like to have graded anti-commutativity or
$\{f,g\}=-(-1)^{|f||g|}\{g,f\}$. So a shift of the degree by 1
solves the problem. This shift can be achieved by considering the
Lie algebra of Hamiltonian vector fields generated by the
observables instead.

If $\omega$ is the symplectic form of the BV space, then the
Hamiltonian vector field generated by a function is defined such
that%
\bea
 {\cal L}_{\BB{X}_f}g \equiv \{f,g\}~,\nn
\eea
where $g$ is any
function on the BV space and $\BB
X_{f}=\iota_{df}\omega^{-1}$. Since $\omega$ has degree $-1$,
$\deg\BB X_f=\deg f+1$.  We have the relation $[\BB X_f,\BB X_g]= \BB X_{\{f,g\}}$, note
the degree shift converts the Gerstenhaber algebra on the right hand side  to
the super Lie algebra on the left hand side.  The Hamiltonian vector fields  $\BB{X}_f$ are in one to one
 correspondence with Hamiltonians $f$ modulo constants. Thus we can fix all functions
  to vanish at a given point to remove this ambiguity.
The Chevalley-Eilenberg (CE) complex of the Lie algebra of such
Hamiltonian vector fields at degree $n$ is spanned by $n$-chain
\bea c_n=\BB{X}_{f_0}\wedge\cdots \wedge \BB{X}_{f_n} ~ . \nn
\eea%
The
boundary operator for such a chain is the conventional one%
\bea\partial \left ( \BB{X}_{f_0} \wedge \BB{X}_ {f_1}, ... ,
\BB{X}_{f_n} \right )&=&\sum_{i<j}\textrm{sgn}_{ij}(-1)^{|f_i|}  \BB{X}_{\{f_i,f_j\}} \wedge
\BB{X}_{f_0}\wedge  ... \wedge
\widehat{\BB{X}}_{f_i}\wedge  ... \wedge \widehat{\BB{X}}_{f_j}\wedge  ... \wedge \BB{X}_{f_n}~,\nn
\eea%
where the sgn is the Koszul sign factor
$(-1)^{(|f_0|+\cdots+|f_{i-1}|)|f_i|+(|f_0|+\cdots
+|f_{j-1}|)|f_j|-|f_i||f_j|}$, which accounts for the minus's
caused by moving $\BB{X}_{f_i}$ and $\BB{X}_{f_j}$ to the front.
Here we make a remark about the convention of graded
(anti)-commutativity. One can either understand
$\BB{X}_{f}\wedge\BB{X}_{g}$ as graded anti-commutative, i.e.
$\BB{X}_{f}\wedge\BB{X}_{g}=-1\times(-1)^{|\BB{X}_f||\BB{X}_g|}\BB{X}_{g}\wedge\BB{X}_{f}$.
Another point of view is to shift the degree $\BB{X}_f$ up by 1
and call it graded commutative:
$\BB{X}_{f}\wedge\BB{X}_{g}=(-1)^{(|\BB{X}_f|+1)(|\BB{X}_g|+1)}\BB{X}_{g}\wedge\BB{X}_{f}$.
The two views make no difference so long as $\BB{X}_f$ has degree
0, yet in working with graded manifolds, the latter is more
advantageous, for then all the commutation relations are
controlled by the degree. In the above Koszul sign, we used the
latter convention, therefore $\deg{\BB{X}_f}=\deg f-1+1$ ($-1$
because the symplectic form has degree $-1$) and
$\BB{X}_f\wedge\BB{X}_g=(-1)^{|f||g|}\BB{X}_g\wedge\BB{X}_f$.

The cochains of the CE complex are just the dual of the chains
$c^n :\ c_n \rightarrow \mathbb{R}$. The differential $\delta$ for
the cochain is induced from $\partial$ through $\delta
c^n(c_{n+1})=c^n(\partial c_{n+1})$.

These definitions fit neatly into the BV framework as follows.
Consider all functions  $f_i$ which  satisfy $\Delta_qf_i=0$ then the the corresponding
 Hamiltonian vector fields $\BB{X}_{f_i}$ give rise to a closed Lie algebra $\BB{A}_q$ since
  $$[\BB X_{f_i},\BB X_{f_j} ]=  (-1)^{|f_i|} \BB X_{\Delta_q (f_i f_j)}~.$$
 We can construct the $n$-chains and boundary operator for $\BB{A}_q$ in the way described above.
Using the property (\ref{bracketquantumus}) we can prove the following identity
\bea\label{lale03030}
 \Delta_q(f_0f_1 ...  f_{n}) = \sum_{i<j}\textrm{sgn}_{ij}(-1)^{|f_i|}~  \{f_i,f_j\}~ f_0~ ...
~\widehat{f_i} ~ ... ~ \widehat{f_j} ~ ... ~ f_n ~.
\eea
In BV context we have a naturally defined cochain, which evaluated  on $\BB{X}_{f_0}\wedge
\BB{X}_{f_1} \wedge ... \wedge \BB{X}_{f_{n}}$ according to the following expression
\bea
c^n(\BB{X}_{f_0}\wedge \BB{X}_{ f_1} \wedge ... \wedge \BB{X}_{f_{n}}) \equiv \int\limits_{\cal L}\
 f_0 ~ f_1~... ~f_{n} ~e^{-S}\in
\mathbb{R}~.\label{path_int_cocyc}\eea%
 One can check easily that it is a multilinear functional with the correct symmetry properties.
This cochain defined through the path integral is in fact a
\emph{cocycle}. This is shown by using the definition of the
coboundary operator and the relation (\ref{lale03030})
\bea
\delta c^n(\BB{X}_{f_0}\wedge \BB{X}_{f_1} \wedge ... \wedge \BB{X}_{f_{n+1}} )&=&c^n\left ( \partial (
\BB{X}_{f_0}\wedge \BB{X}_{f_1} \wedge ... \wedge \BB{X}_{f_{n+1}} )\right )\nn \\
 = \int\limits_{\cal L}
\Delta_q(f_0~f_1~...~ f_{n+1})~e^{-S}
&=& \int\limits_{\cal L} \Delta(f_0~f_1~ ... ~ f_{n+1}~e^{-S})=0~,\nn\eea%
where in the last step we used the fact the integral of any
$\Delta$-exact function is zero.

We would like to emphasize that the cochain thus defined
\emph{does depend on the choice of the Lagrangian submanifold}.
Although each $f_i$ obeys $\Delta_q(f_i)=0$, $\Delta_q(f_0\cdots
f_{n})\neq0$ in general, so the Stokes theorem does not apply.
Hence we denote the cochain by $c^n_{\cal L}$ and we study the
${\cal L}$ dependence next. By Schwarz's explicit construction,
every ${\cal L}$ is locally embedded in the BV space as
$T^*[-1]M=T^*[-1]{\cal L}$; the simplest ${\cal L}$ namely $M$
itself is such an example. If we denote the coordinates of ${\cal
L}$ as $x^a$ and $x^+_a$ that of the transverse direction to
${\cal L}$ (${\cal L}$ is given by $x^+=0$ locally). Then any
small
deformation is parameterized as%
\bea x_a^+=\frac{\partial}{\partial x^a}\Psi(x)~.\nn\eea%
The function $\Psi$ only depends on $x$ and may be regarded as the
generating function for the canonical transformation going from
${\cal L}$ to ${\cal L}+\delta{\cal L}$. Locally, the Laplacian is
$\Delta=\partial_{x^a}\partial_{x^+_a}$, so $\Delta \Psi=0$
trivially.
Now
\bea
(\int\limits_{{\cal L}+\delta{\cal L}}-\int\limits_{{\cal
L}})f_0~f_1~...~ f_n ~ e^{-S}&=&\int\limits_{{\cal
L}}\Psi~\frac{\overleftarrow{\partial}}{\partial
x^a}\frac{\overrightarrow{\partial}}{\partial
x^+_a}\big(f_0~f_1~...~ f_n ~ e^{-S}\big)=\int\limits_{{\cal
L}}\{\Psi,f_0f_1... f_n e^{-S}\}\nn\\%
&=&-\int\limits_{{\cal L}}\big(\Delta(\Psi~ f_0~f_1~...~f_n ~
e^{-S})+\Psi~\Delta(f_0~f_1~...~ f_n ~ e^{-S})\big)\nn\\%
&=&-\int\limits_{{\cal L}}\Psi~\Delta_q(f_0~f_1~...~ f_n)~e^{-S}~.\nn
\eea%
If we define a new $(n-1)$-cochain by
\bea
\tilde{c}^{n-1}(\BB{X}_{f_0}\wedge \BB{X}_{f_1}\wedge ... \wedge \BB{X}_{f_{n-1}})\equiv -\int_{{\cal
L}}\Psi~(f_0~f_1~... ~ f_{n-1})~e^{-S}.\nn\eea%
This cochain is \emph{not} closed, however we have%
\bea
(c^n_{{\cal L}+\delta{\cal L}}-c^n_{{\cal L}})(\BB{X}_{f_0}\wedge \BB{X}_{f_1}\wedge ... \wedge
\BB{X}_{f_{n}})&=&\tilde c^{n-1}_{{\cal L}}\left (\partial (\BB{X}_{f_0}\wedge \BB{X}_{f_1}\wedge ...
\wedge \BB{X}_{f_{n}}) \right )\\
&=&
\delta\tilde c^{n-1}(\BB{X}_{f_0}\wedge \BB{X}_{f_1}\wedge ... \wedge \BB{X}_{f_{n}})~.\label{coboundary}
\eea%
Our observation is thus, the change of the LagSubMfld changes the
cochain $c^n$ by a coboundary $\delta\tilde c^{n-1}$. Thus for any
choice of ${\cal L}$, the path integral gives a representative of
the class in the cohomology of the Lie algebra of the quantum
observables. Yet two choices of ${\cal L}$ that are not homotopic
to each other will produce different classes in the cohomology.

So far our discussion has been formal, and may only be applied
properly to a finite dimensional BV manifold. While for most cases
of interest to physics, the BV space is the space of mappings and
hence of infinite dimension. One usually does not have a well
defined Laplacian, and the condition $\Delta_q f=0$ can at best be
realized formally. Another drawback is that the relevant Lie
algebra cohomology is on the space of mappings, while we quite
often would like to ask questions about the properties of the
target manifold alone, the formalism developed above becomes
unwieldy. In the section \ref{sec_A_3D_Model} we will set up a 3D
topological field theory that focuses on the Lie algebra
cohomology of Hamiltonian vector fields of the target manifold.
The discussion there is along the lines of
 \cite{Hamilton:2007tg}.

But before we do so, we would have to digress a little for some other
background material.

\section{Background material}
\label{sec_review}

In this section, we review the relevant background material.
 We remind the idea behind the construction of characteristic classes
  of flat bundles.   We hint on the application of this construction
   within BV formalism.
We also review the necessary  facts concerning Lie
algebra homology of formal Hamiltonian vector fields and its
relation to the graph homology.

\subsection{Characteristic Classes for Flat Bundles}
\label{sub-charclass}
Consider the principal bundle $P$ over base $M$  with structure
group ${\mathbf G}$,
\begin{equation}
\begin{CD}
 P @  << <  {\mathbf G}    \\
  @ V  VV\\
 M
\end{CD}
\end{equation}
 If we choose the connection $A$ on $P$ with curvature $R$, then $R$
 is a Lie algebra valued $2$-form on $M$.   The procedure
we are familiar with is to take an invariant polynomial of the
generators of the Lie algebra ${\mathbf g}$ (usually a trace or a
determinant), and plug in the curvature 2-form $R$. The Chern-Weil
theorem guarantees that the resulting form is a closed form and so
we have the mapping%
\bea \mathbb{C}[{\mathbf g}^*]^{Ad_{\mathbf G}}\rightarrow
H^{2k}(M,\mathbb{R})~.\nn\eea
 This is the standard construction of the classical characteristic classes for the principle
  bundles.

 A flat bundle is a principal bundle equipped with a connection whose curvature vanishes identically,
  flat connection.  Thus, by the Chern-Weil theory all characteristic classes vanish and it may
   appear that the flat bundle is close to a trivial bundle. However, it is far from being true.
   Let us sketch the main idea behind the construction of the characteristic classes for flat bundles,
    which are also called secondary characteristic classes.
Now we use the connection rather
than the curvature. For the Lie algebra ${\mathbf g}$ there is the CE
cochain complex $c^{\bullet}= \wedge^{\bullet}{\mathbf g}^*$ with the standard
CE differential. Instead of invariant polynomials, take any
cocycle $c^q$ in this complex and plug in the connection,
resulting in a differential form on the bundle $P$ given by%
\bea c^n~\stackrel{A}{\longrightarrow}~c^n(\underbrace{A,~...,~
A}_{n+1}) \in \Omega^{n +1}(P)~.\label{Ch_W}\eea%
This mapping from the cochain complex to the differential forms on $P$  does not yet
 send cochain differential to de Rham
differential. To mend this, one must require the connection to be
flat, i.e. it satisfies the Maurer-Cartan equation $dA+A\wedge
A=0$. To make it look more familiar, we pick a basis $t^a$ for  the Lie algebra ${\mathbf g}$
 and we can write the
flatness condition as%
\bea
 (dA_a)(t^a)+\frac{1}{2}(A_b\wedge A_c)[t^b,t^c] =0~, \nn\eea
  where $[~,~]$ is Lie bracket for ${\mathbf g}$.
The last identity makes it clear that the flatness condition
qualifies the mapping (\ref{Ch_W}) as a differential graded map, for%
\bea
&&dc^n(A, ~... ,~A)=d(A_{a_0}\wedge ... \wedge A_{a_n})~c^n(t^{a_0},\cdots,t^{a_n})\nn\\%
&=&-\frac{1}{2}\sum_i(-1)^i A_{a_0}\wedge ... \wedge  \underbrace{A_b\wedge
A_c}_{i}\wedge... \wedge
A_{a_n}c^n(t^{a_0},\cdots,\underbrace{[t^b,t^c]}_{i},t^{a_n})\nn\\%
&=&-\frac{1}{2}A_{a_0}\wedge ... \wedge  A_{a_{n+1}}(\delta
c^n)(t^{a_0},\cdots,t^{a_{n+1}})~.\label{ismow92929} \eea
Moreover, if $c^n$ is a cocycle in the CE complex, the map
(\ref{Ch_W}) gives us a closed form on $P$.
 Thus the flat connection induces the map of the cohomology
 groups
 \beq\label{homo123}
  H^{n} ({\mathbf g}, \mathbb{R})~\stackrel{A}{\longrightarrow}~ H^{n+1} (P, \mathbb{R})
  ~\stackrel{s}{\longrightarrow}~ H^{n+1} (M, \mathbb{R})~,
 \eeq
  where the last step involves the choice of the section $s$ (or trivialization of $P$).
    The above map does not change if we choose another trivialization of $P$ in
     the same homotopy  class of trivializations.   This is the construction of the
secondary characteristic classes.  This theory can be applied to the case of infinite dimensional
 algebras (groups) as well and it plays the central role in the characteristic classes of foliations.
  For further details about the characteristic classes of the flat bundles the
   reader may consult the book by Morita \cite{morita}.

 The flat connections appear a lot in physics. Let us discuss
  the relevant setup in which we generalize
 this slightly to include not just the Lie
algebra valued differential forms but a general $Q$-structure.
Recall a $Q$-structure is a degree one vector field satisfying
$Q^2=0$. As a $Q$-structure is a natural generalization of the de
Rham differential, the $Q$-equivariant fiber bundles are the
generalization of flat bundles in the following way. Given any
fiber bundle ${\cal E}\stackrel{\pi}{\rightarrow} {\cal M}$,
suppose there is $Q$ structure over graded manifold ${\cal M}$ and
a $\tilde Q$ over total space ${\cal E}$,
 which is also graded manifold.  The $Q$-equivariantness says $\pi_{*}\tilde Q=Q$. In a
   local coordinate
such $\tilde Q$ can be written as (taking $e^I$ as the coordinates
of
the fiber)
\bea \tilde Q(x, e)=Q(x) +A^I (x, e) \frac{\partial}{\partial
e^I}~,\nn\eea%
where $A^I$ is a vector field along a fiber. $\tilde Q^2=0$ implies that
$A$ satisfies the Cartan-Maurer equation%
\bea\label{2829sskska}
QA+\frac{1}{2}[A,A]=0\label{Qequiv}~,\eea%
 where $[~,~]$  stands for the Lie bracket of vector fields along the fiber.
  Thus in this setup the Lie algebra ${\mathbf g}$ can be identified with the algebra of  formal
   vector fields along the fiber.
  By using the construction analogous to (\ref{ismow92929})
 one obtains $Q$-closed functions by evaluating the $A$ on the cocycle
  of this infinite dimensional algebra of ${\mathbf g}$.
  These $Q$-closed functions are the
characteristic classes for the $Q$-structure. As the $Q$-structure
includes a wide variety of differentials such as the de Rham,
Doubeault, Chevalley-Eilenberg, Poisson-Lichnerowicz etc we have a
more uniform way of investigating the characteristic classes
associated with these structures.

There is an immediate application of these ideas in the BV path
integral framework. Recall from section \ref{sec_BV} that
\beq
 c^n (\BB{X}_{f_0} \wedge ... \wedge \BB{X}_{f_n}) = \int\limits_{\cal L} f_0 ~... ~f_n
\eeq
 defines the cocycle for the Lie algebra of divergenceless Hamiltonian vector fields (i.e., $\Delta f_i =0$)
  on BV space.  Consider the BV action $S$ which satisfies $\Delta S =0$.
Suppose that the action also depends on some extra parameters and
that there exists another odd differential $Q$ acting on
those parameters, such that
\bea Q S+\frac{1}{2}\{S,S\}=0~.\label{Cartan_Maurer}
\eea
 This is a quite typical setup in BV theory. Equation (\ref{Cartan_Maurer}) appears as a consequence
  of the classical master equation and the extra parameters can originate from the zero modes of the theory, for example.
   Now let us evaluate the partition function of this BV theory
\beq
 Z = \int\limits_{\cal L} e^{-S} = \sum\limits_{n=0}^\infty \frac{(-1)^n}{n!} c^{n-1} (\BB{X}_{S} \wedge ... \wedge \BB{X}_S)~,
\eeq
where $c^n (\BB{X}_{S} \wedge ... \wedge \BB{X}_S)$ is a cocycle since $\Delta S=0$ and it is now a function
of the extra parameters.  We can show easily that this function is
annihilated by $Q$
\bea
Q c^n (\BB{X}_{S} \wedge ... \wedge \BB{X}_S)= -\frac{1}{2}
c^{n}(\partial(\BB{X}_S\wedge ... \BB{X}_S)\big)=0~,\label{general_CM}\eea
 where we used the property (\ref{Cartan_Maurer}).
The most important example where this situation can arise is of
course when we have a bundle structure whose fiber is equipped
with an odd symplectic form and the extra parameter is the
coordinate of the base. Then the relation  (\ref{Cartan_Maurer})
is nothing but the $Q$-equivariantness condition (\ref{Qequiv}),
namely the $Q$-structure  on the base is lifted to $\tilde
Q=Q+\{S,\cdot\}$ in the total space. Within this picture the
partition function $Z$
 gives rise to $Q$ characteristic class (the concrete representative depends on
  the choice of ${\cal L}$).  Although the present argument is formal, we will argue later
   that this is a generic feature of 3D TFTs.

\subsection{Lie Algebra/Graph Cohomology}
In this subsection we review briefly the algebra of formal Hamiltonian vector fields.
 We will use these material in the next sections.

Consider the vector space $\mathbb{R}^{2m}$ equipped with the canonical symplectic structure.
Let $\mathbf{Ham}^0_{2m}$ be the Lie algebra of formal (polynomial) Hamiltonian vector fields
over $\BB{R}^{2m}$ preserving the origin; let $\mathbf{Ham}^1_{2m}$ consist
 of those elements of $\mathbf{Ham}^0_{2m}$ whose Taylor expansion
starts from the quadratic term and finally $sp(2m, \mathbb{R})$
are those elements whose coefficients are linear. If one chooses to
talk about the Hamiltonian function instead, then $sp(2m,
\mathbb{R})$ corresponds to quadratic polynomials,
$\mathbf{Ham}^1_{2m}$ corresponds to cubic or higher polynomials.
Let $C_{\bullet}(\mathbf{Ham}^0_{2m})$ be the Chevalley-Eilenberg
complex of $\mathbf{Ham}^0_{2m}$ and $sp(2m, \mathbb{R})$ acts on
this complex through the adjoint action. We shall consider the
$sp(2m, \mathbb{R})$ coinvariants\footnote{The coinvariants, in
contrast to the invariants, are the largest quotient of
$C_{\bullet}(\mathbf{Ham}^0_{2m})$ on which $sp(2m, \mathbb{R})$
acts trivially, or simply speaking, the orbits of the $sp(2m,
\mathbb{R})$ action.} of the complex
$C_{\bullet}(\mathbf{Ham}^1_{2m})$. If we denote such coinvariants
as $C_{\bullet}(\mathbf{Ham}^0_{2m}, sp(2m, \mathbb{R}))$, then we
have the isomorphism due to
Kontsevich \cite{Kontsevich:formal} that%
\bea H_{\bullet}(\mathbf{Ham}^0_{2m}, sp(2m, \mathbb{R}))\sim H_{\bullet}({\cal
G})\label{iso_graph_Lie}~,\eea%
where ${\cal G}$ is the (undecorated) graph complex. The reason
for 'modding' out the $sp(2m, \mathbb{R})$ subgroup will become
clear once we consider this isomorphism from the path integral
point of view. The same isomorphism (\ref{iso_graph_Lie}) can be
generalized to the superspace $\mathbb{R}^{2m|k}$ with the even
symplectic structure, see \cite{hamilton-2006}.

 We use here the same conventions as in the previous section.
  However we are interested in a different Lie algebra now.
We use $\BB{X}_f$ to denote a Hamiltonian vector field  generated by $f$ over
$\mathbb{R}^{2m}$ with the canonical symplectic structure.
 The CE complex will be spanned
by
the exterior product of the form%
\bea c_n=\BB{X}_{f_0}\wedge\cdots \wedge \BB{X}_{f_n}~.\nn\eea%
The Chevalley-Eilenberg boundary operator is %
\bea \partial
c_n=\sum_{i<j}(-1)^{i+j+1}[\BB{X}_{f_i},\BB{X}_{f_j}]\wedge\BB{X}_{f_0}\cdots
\wedge\widehat{\BB{X}}_{f_i}\cdots
\wedge\widehat{\BB{X}}_{f_j}\cdots\wedge\BB{X}_{f_n}~.\nn\eea%
By using the relation
$[\BB{X}_{f},\BB{X}_{g}]= \BB{X}_{\{f,g\}}$, we can abbreviate%
\bea
 \BB{X}_{f_0}\wedge\cdots \wedge \BB{X}_{f_n} \textrm{ as }
(f_0,\cdots ,f_n)~,\label{abbre}\eea%
and the boundary operator by \bea \partial(f_0,\cdots
f_n)=\sum_{i<j}(-1)^{i+j+1}(\{f_i,f_j\},f_0,\cdots
\hat{f_i},\cdots\hat{f_j},\cdots f_n)~.\label{Lie_diff_even}\eea%

Apart from the petit details, the mapping in (\ref{iso_graph_Lie})
is easy to understand. Take the Euclidean space $\mathbb{R}^{2m}$
equipped with the standard symplectic structure
$\sum\limits_{\mu<\nu}\Omega_{\mu\nu}dx^\mu \wedge dx^\nu$. The
function $f$'s are all polynomials on $\mathbb{R}^{2m}$, so a
given chain corresponds to a sum%
\bea c_n=\sum (\mathfrak{m}_0, \mathfrak{m}_1,\cdots, \mathfrak{m}_n)~,\nn\eea%
where $\mathfrak{m}_i$ are all monomials. An $l$-th order monomial will
correspond to an $l$-valent vertex in the graph. For every
propagator connecting leg $\mu$ and $\nu$ one incorporates a factor
$\Omega_{\mu\nu}$ into the coefficient of the graph
\bea (x^\mu x^\nu x^\rho)\sim\Diagram{\vertexlabel^\mu& &\\ fd f \vertexlabel^\rho\\
fu \\\vertexlabel_\nu},\ \  (x^\mu x^\nu x^\rho x^\lambda)\sim\Diagram{\vertexlabel^\mu& &\vertexlabel^\lambda\\ fd fu\\ fu fd\\
\vertexlabel_\nu& &\vertexlabel_\rho},\ \
\Omega_{\sigma\gamma}\big((x^\mu x^\nu x^\sigma),(x^\rho x^\lambda x^\gamma)\big)
\sim\Diagram{\vertexlabel^\mu&\vertexlabel_\sigma&\vertexlabel_\gamma&\vertexlabel^\lambda\\ fd & & fu\\ &f&\\
fu& & fd\\ \vertexlabel_\nu& &
&\vertexlabel_\rho}\label{LC_GC_Correspondence}\eea

The Poisson bracket between two monomials of degree $p$ and $q$ produces
a sum of monomials of order $p+q-2$, so%
\bea
&&\partial\big(\Omega_{\sigma\gamma}(\cdots,(x^\mu x^\nu x^\sigma),(x^\gamma x^\rho x^\lambda ),\cdots)\big)
=\cdots+\Omega_{\sigma\gamma}(\cdots,\{x^\mu x^\nu x^\sigma ,x^\gamma x^\rho x^\lambda\},\cdots)+\cdots\nn\\%
&=&\cdots+(\cdots,x^\mu x^\nu x^\rho x^\lambda,\cdots)+\cdots,\nn\eea%
where we have only focused on the propagator $\sigma\gamma$ while assuming
the legs $\mu \nu \rho \lambda$ are connected to other parts of the graph in a
certain way.
In the graph language the boundary operator acts as%
\bea \partial\Diagram{\vertexlabel^\mu &fd & & fu
\vertexlabel^\lambda \\\vertexlabel_\nu & &fv&\\ &
fu& & fd &\vertexlabel^\rho}=\pm\Diagram{\vertexlabel^\mu& &\vertexlabel^\lambda\\ fd fu\\ fu fd\\
\vertexlabel_\nu& &\vertexlabel_\rho};\hspace{.5cm}
\partial\Diagram{\vertexlabel^\mu & & &\vertexlabel^\lambda\\ fd & & fu\\ &f&\\
fu& & fd\\ \vertexlabel_\nu& & &\vertexlabel_\rho}=\pm\Diagram{\vertexlabel^\mu& &\vertexlabel^\lambda\\ fd fu\\ fu fd\\
\vertexlabel_\nu& &\vertexlabel_\rho}~.\nn\eea%
So the boundary operator acts on a graph by deleting one
propagator. This is exactly the differential for the graph
complex. We have omitted lots of details, especially those
concerning how to work out the sign factors and the orientation of
the graph; the reader may see \cite{hamilton-2006} for a
 full treatment. The similar construction can be applied to the superspace ${\mathbb R}^{2m|k}$ with even symplectic
  structure.
  In what follows we use the Greek letters for even case ${\mathbb R}^{2m}$, while upper case Latin letters for the supercase
   ${\mathbb R}^{2m|k}$.

\section{3D  AKSZ  topological field theory}\label{sec_A_3D_Model}

The Alexandrov-Kontsevich-Schwarz-Zaboronsky (AKSZ) construction \cite{Alexandrov:1995kv}
allows one to produce a variety of topological $\sigma$-models in a rather  canonical fashion.
 The AKSZ approach uses mapping space between a source supermanifold (graded manifold)
  and target supermanifold (graded manifold). The source and target manifolds are  equipped with additional
structures.
    In this section we define 3D AKSZ model and
explore the geometrical meaning of the correlation functions in
AKSZ model on a 3D source manifold, the result turns out related
to the Lie algebra cohomology of Hamiltonian vector fields on the
\emph{target space}.  In particular we make some remarks on the perturbative theory.
The correlators of the two 3D AKSZ models we
know, namely the Chern-Simons theory and the Rozansky-Witten
model, both fit this description.

\subsection{Construction of AKSZ model}

The general construction of the AKSZ
models is standard by now. We shall be brief and
 concentrate our attention only on 3D models.
For more details the reader may consult
\cite{Alexandrov:1995kv, Cattaneo:2001ys, Roytenberg:2006qz}.

 Consider an \emph{even symplectic} supermanifold ${\cal M}$ with the local coordinates
  $X^A$ and  $|A|$ denotes the degree $X^A$.  Suppose the even symplectic form on the
target manifold is $\Omega$, and it can be written locally as the
differential of the Liouville form $\Omega=d\Xi$. Assume we are in
a Darboux coordinate then $\Xi=X^A~\Omega_{AB}~dX^B$. We will refer to ${\cal M}$
 as a target. Also let us consider the three dimensional manifold $\Sigma_3$.
  We are interested in the odd tangent bundle $T[1]\Sigma_3$,
where $\xi$ is the
bosonic coordinate of $\Sigma_3$ and $\theta$ is the odd fiber
coordinate of $T[1]\Sigma_3$. We will refer to $\Sigma_3$ as source.
 In the following discussion
 we have chosen
the source manifold $\Sigma_3$ to be $S^3$ or more generally a
rational homology sphere.

Consider  the mapping space Maps$(T[1]\Sigma_3, {\cal M})$ and
 denote the mapping by $\BS{X}^A(\xi,\theta)$.
The even symplectic form on ${\cal M}$  induces an odd symplectic form
in the space of mappings Maps$(T[1]\Sigma_3,{\cal M})$ according to%
\bea \omega=\frac{1}{2}\int\limits_{T[1]\Sigma_3} d^6z~\big
(\Omega_{AB}\delta\BS{X}^A\delta\BS{X}^B\big)~,\label{symsp22929}\eea%
where we write $\xi,\theta$ collectively as $z$ and $d^6z\equiv d^3\theta
~d^3 \xi$.
Note that each $\BS X$ is a superfield, hence can be expanded
into components%
\bea
\BS{X}(\xi,\theta)=X(\xi)+\theta^aX(\xi)_a+\frac{1}{2}\theta^b\theta^aX(\xi)_{ab}
+\frac{1}{3!}\theta^c\theta^b\theta^aX(\xi)_{abc}~.\nn\eea%
The components correspond to forms of different degrees on $\Sigma_3$.
When we do not want to spell out all the indices, we will write%
\bea X_{(0)}=X(\xi)~;\ X_{(1)}={X}(\xi)_ad\xi^a~;\
{X}_{(2)}=\sum_{a<b} {X}(\xi)_{ab}d\xi^a\wedge d\xi^b~; ~~...
\nn\eea We may obtain the odd symplectic form written in
components by
integrating out $d^3\theta$ in (\ref{symsp22929}) (details left to the appendix)%
\bea \omega=-\int\limits_{\Sigma_3} d^3\xi~
\Omega_{AB}\big(\delta{X}^A_{(3)}\wedge\delta{X}^B_{(0)}
+\delta{X}^A_{(1)}\wedge\delta{X}^B_{(2)}\big)~.\label{sympe2929201}\eea%
The BV space Maps$(T[1]\Sigma_3,{\cal M})$  is infinite dimensional and there is no well defined
measure for the path integral, we shall use the naive one%
\bea \textrm{vol}=\wedge^{top}d{X}_{(0)}\wedge^{top}d{X}_{(1)}
\wedge^{top}d{X}_{(2)}\wedge^{top}d{X}_{(3)}~.\nn \eea
 With this
volume form we have a naive odd Laplacian \bea \Delta \equiv
\int\limits_{\Sigma_3}
d^3\xi~(\Omega^{-1})^{AB}(-1)^{|A||B|}\big(\frac{\delta}{\delta{X}^A_{(3)}(\xi)}\frac{\delta}{\delta{X}^B_{(0)}(\xi)}
+\frac{\delta}{\delta{X}^A_{(1)}(\xi)}\frac{\delta}{\delta{X}^B_{(2)}(\xi)}\big)~.\nn
\eea Note that the odd Laplacian is the restriction of a
distribution on $\Sigma_3 \times \Sigma_3$ to the diagonal, and is
a singular object in this infinite dimensional context. Thus it
should be understood as the limit of a suitably regularized
expression (see appendix). The odd Laplacian has a number of the
formal properties
\bea &&\Delta\int\limits_{T[1]\Sigma_3} d^6z\ f(\BS X(z))=0~,\nn\\%
&&\Delta\big(\int\limits_{T[1]\Sigma_3} d^6z_1f(\BS
X(z_1))\int\limits_{T[1]\Sigma_3}d^6z_2\ g(\BS
X(z_2))\big)=(-1)^f\int\limits_{T[1]\Sigma_3}
d^6z\ \{f(\BS X(z)),g(\BS X(z))\}~,\nn \\%
&&\{\int\limits_{T[1]\Sigma_3} d^6z\ f(\BS X(z)),\int\limits_{T[1]\Sigma_3}
d^6z\ g(\BS X(z))\}=-\int\limits_{T[1]\Sigma_3} d^6z\
\{f(\BS X(z)),g(\BS X(z))\}~.\label{formal} \eea
 We refer the reader to the appendix for further details.

 Let us choose an odd function $\Theta$  on ${\cal M}$ which satisfies $\{ \Theta, \Theta \}=0$ with
  respect to the even symplectic structure on ${\cal M}$.  Then
the AKSZ construction gives the standard BV action%
\bea &&S=S_{kin}+S_{int}=\int\limits_{T[1]\Sigma_3} d^6z\ \BS{X}^A \Omega_{AB} D{\BS X}^B
 +\BS{{X}}^*\Theta~,\label{AKSZ_action}\\%
&&D\equiv \theta^a\partial_a~,\nn\eea%
where the first term involving the Liouville form is called the
kinetic term. The kinetic term is independent of the concrete
choice of the Liouville form. $\BS{{X}}^*\Theta$ is pullback of $\Theta$ through
$\BS{{X}}$ to the space of mappings. It serves as the interaction
term. We often write the pull back $\BS{X}^*\Theta$ simply as
$\BS\Theta$. One can check easily that $S$ satisfies the classical master equation $\{S, S\}=0$
 with respect to
\bea
\{S,S\}=-\int\limits_{T[1]\Sigma_3} d^6z\
D(\BS{\Xi}_AD\BS{X}^A+\BS\Theta)+ \BS{X}^*\{ \Theta, \Theta\}=-\int\limits_{T[1]\Sigma_3}
d^6z\ \{\BS\Theta,\BS\Theta\}=0~,\nn\eea%
where the first term drops because it is a total derivative and
$\Sigma_3$ has no boundary. So the only requirement is merely
$\{\Theta,\Theta\}=0$. In the expression above and in further
discussion we use the same notations for
 the bracket on ${\cal M}$ and for the BV bracket on Maps$(T[1]\Sigma_3,{\cal M})$.
  Hopefully it is not confusing since it can be understood from the context which bracket is used.
   Thus the action (\ref{AKSZ_action}) formally satisfies the quantum master equation $\Delta e^{-S}=0$, by using
the first property of $\Delta$ in (\ref{formal}) and  $\{S,S\}=0$,
 For further reference the component form of  kinetic term and the interaction terms
 are
\bea
 S_{kin}&=&\int\limits_{\Sigma_3}
d^3\xi~\Omega_{AB}\big(-{X}^A_{(1)} \wedge d{X}^B_{(1)}+{X}^A_{(2)}\wedge
d{X}^B_{(0)}\big)~,\label{kinetic_comp}\\
S_{int} &=& \int\limits_{\Sigma_3} d^3\xi~ X_{(3)}^A  ~(\partial_A \Theta) +X_{(2)}^A \wedge X_{(1)}^B~ ( \partial_B \partial_A \Theta)
 + \frac{1}{6} X_{(1)}^A \wedge X_{(1)}^B \wedge X_{(1)}^C~( \partial_C \partial_B \partial_A \Theta)~.\nn
\eea

  In the present discussion we keep in mind  only  $\mathbb{Z}_2$-grading. The construction can be
   refined to $\mathbb{Z}$-grading with the source and target being graded manifolds equipped with
    the extra structure \cite{Roytenberg:2006qz}, the main example is given by the Courant sigma model. The
     Chern-Simons  theory is special case of the Courant sigma model with the target being ${\cal M} ={\mathbf g}[1]$,
      where ${\mathbf g}$ is a metric Lie algebra.
    In principle we will allow the BV theory to depend on extra free
      parameters, e.g. $\Theta$ may depend on the parameters
     other then the coordinates of ${\cal M}$.

\subsection{Formal Properties of Correlators}
\label{sub-formcor1}

  In this subsection we apply the formal observation from  section \ref{sec_BV} to 3D AKSZ theory
   constructed in the previous subsection.  The simple observation is that certain subalgebra of
    quantum observables can be mapped to specific  subalgebra of Hamiltonian vector fields on ${\cal M}$.
   Thus the corresponding correlator can be interpreted entirely in term of
\emph{target} space geometry.


Consider the objects of the form $\BS{F} \equiv\int d^6z\ f(\BS X(z))$. According to the property (\ref{formal})
 the BV bracket between $\{\BS{F}_i, \BS{F}_j\}$  gets mapped to the even bracket $\{f_i, f_j\}$ on ${\cal M}$. Moreover
 $\BS{F}$'s are quantum  observables
if they satisfy \bea 0=\Delta\big(e^{-S}\int\limits_{T[1]\Sigma_3}
d^6z\ f(\BS X(z))\big)=e^{-S}\int\limits_{T[1]\Sigma_3} d^6z\
Df(\BS{X}(z))+\BS{X}^*\{\Theta,f(X)\}~.\nn\eea
 Thus the only requirement on $\BS{F}$ to be a quantum observable is that
  the corresponding $f$ commutes with $\Theta$ on ${\cal M}$, i.e. $\{\Theta, f\}=0$.
The observables of this type form a closed algebra and we define
the correlator using the path integral \bea \langle \BS{F}_0
\BS{F}_1 ... \BS{F}_n \rangle  \equiv \int\limits_{\cal L}
\big(\Large{\textrm{$\smallint$}}d^6z_0f_0(z_0)\Large{\textrm{$\smallint$}}d^6z_1f_1(z_1)\cdots
\Large{\textrm{$\smallint$}}d^6z_n
f_n(z_n)\big)e^{-S},\label{requ}\eea%
where $f(z)$ is the short hand of $f(\BS X(z))$. Repeating the
formal argument
 from section \ref{sec_BV} and using the fact that $\BS{F}$'s algebra is mapped to
  $f$'s algebra on ${\cal M}$ we arrive at the conclusion that the correlator
\bea \langle \BS{F}_0 \BS{F}_1 ... \BS{F}_n \rangle =
c^n(\BB{X}_{f_0}\wedge \BB{X}_{ f_1} \wedge ... \wedge
\BB{X}_{f_{n}}) \eea
 corresponds to cocycle of the Lie algebra of Hamiltonian vector fields $\BB{X}_{f}$ which commute with $\BB{X}_\Theta$.
 We once again remark that this way of writing the correlation function (\ref{requ})
agrees with the graded commutativity on the left hand side, namely\footnote{To avoid ugly expressions we adopt
 the simple notation for degree, namely $\deg f =|f|= f$. Since the degree is essential only for signs, it appear
  in the expressions like $(-1)^f$  and thus there should be no confusion.}
$\BB{X}_{f}\wedge\BB{X}_{g}=+(-1)^{(f-2+1)(g-2+1)}\BB{X}_{g}\wedge\BB{X}_{f}$
($-2$ because the symplectic form has degree 2). Because the
integration measure $d^6z$ carries $-3$ degree, so%
\bea
\int\limits_{T[1]\Sigma_3}d^6z_1f(z_1)\int\limits_{T[1]\Sigma_3}
d^6z_2g(z_2)=(-1)^{(f-3)(g-3)}\int\limits_{T[1]\Sigma_3}d^6z_1g(z_1)\int\limits_{T[1]\Sigma_3}
d^6z_2f(z_2)~.\nn\eea%
One may compare this to the situation of section \ref{sec_BV} the
degree shift is due to the odd symplectic form while here it is
due to the degree of the measure for the source.  Moreover using
 the standard BV manipulations and the correspondence of BV algebra of $\BS{F}$'s with
  the algebra on ${\cal M}$ one can easily check that our correlator is a cocycle
\bea &&\delta  c^{n-1} (\BB{X}_{f_0}\wedge \BB{X}_{ f_1} \wedge ...
\wedge \BB{X}_{f_{n}}) =\int\limits_{\cal
L}\Delta_q\big(\BS{F}_0\BS{F}_1\cdots\BS{F}_n\big)e^{-S}\nn\\%
&=&\int\limits_{\cal L}\sum_{i<j}(-1)^{f_i+ (f_i+3)(f_0+\cdots
f_{i-1}+3i)+(f_j+3)(f_0+\cdots f_{j-1}+3j)+(f_i+3)(f_j+3)} ~\times \nn \\
&&(-1)^n\{\BS{F}_i ,\BS{F}_j \}~
\BS{F}_0\cdots\widehat{\BS{F}_i}\cdots\widehat
{\BS{F}_j}\cdots\BS{F}_ne^{-S} =c^{n-1}(\partial(
\BB{X}_{f_0}\wedge \BB{X}_{ f_1} \wedge ... \wedge \BB{X}_{f_{n}})
)~,\nn\eea
 where everything matches including the signs.

\subsection{Formal Properties of Perturbation Theory}
\label{sub-formcor2}

Our argument so far was quite formal and we would like to convert it into the concrete calculation
 with the precise properties. For this we will have to resolve to the perturbation theory.  Before defining
  the precise Feynman rules, let us make a few comments about the expected properties of the correlators in
   the perturbative theory.

 We can repeat the formal argument from the previous subsection with $\Theta =0$. The correlators are now
\bea
  \int\limits_{\cal L}
\big(\Large{\textrm{$\smallint$}}
d^6z_0f_0(z_0)\cdots\Large{\textrm{$\smallint$}}d^6z_n
f_n (z_n)\big)e^{-S_{kin}}=c^n( \BB{X}_{f_0}\wedge \BB{X}_{ f_1} \wedge ... \wedge \BB{X}_{f_{n}})~, \label{requ339993}\eea%
 which should be cocycles for the Lie algebra of Hamiltonian vector fields on ${\cal M}$.
We can give some general remarks about the structure of the
correlation function. Assuming that $X^A$ are the Darboux
coordinates of the target space, then the kinetic term is
$\BS{{X}}^A\Omega_{AB}D\BS{{X}}^B$. If this
gives a non-degenerate quadratic term when restricted to ${\cal
L}$, then we can invert it and obtain the propagator. In the model
we have the propagator will basically consist of $\Omega^{-1}$ and
the inversion of the de Rham operator $G(z_1,z_2)$ (inverting the
de Rham operator requires one to choose ${\cal L}$ carefully, more
of this later). Applying the Wick theorem the correlation function (\ref{requ339993})
 is represented by the Feynman diagrams (graphs), and the end result is written
schematically as \bea c^n( \BB{X}_{f_0}\wedge \BB{X}_{ f_1} \wedge
... \wedge \BB{X}_{f_{n}})= \langle \BS{F}_0 \BS{F}_1 ...
\BS{F}_n\rangle=\sum_{\Gamma}b_{\Gamma}I_{\Gamma}~,\label{BgammaIgamma}\eea
where we sum over the  graphs $\Gamma$.
 Concretely, any graph $\Gamma$ gives
a particular way of routing the propagators. Since every insertion
$\BS{F}_i$ is integrated with the measure $d^6z_i$, we have an
integration of these propagators over the configuration space
$T[1]\Sigma_3\times\cdots \times T[1]\Sigma_3$. This integral thus
associates a graph with a number which is called the \emph{weight
function} $b_{\Gamma}$.  While $I_{\Gamma}$ corresponds to the
combination of derivatives
 of $f$'s (vertices) contracted by $\Omega^{-1}$ (edges) in the way prescribed by $\Gamma$.
  The essential property of  sum $\sum_{\Gamma}b_{\Gamma}I_{\Gamma}$ is that it should give a cocycle
   for the Hamiltonian vector fields on ${\cal M}$. In the next section we are going to discuss
    the concrete prescription behind the formula (\ref{BgammaIgamma}).

\section{Perturbative expansion of the AKSZ Model}\label{sec_gauge_fix}

In this section we construct the perturbation theory for 3D AKSZ model constructed in the previous section.
 From now on we assume that ${\cal M}$ is super(graded) vector space $\mathbb{R}^{2m|k}$ equipped with
  the canonical even symplectic form.  This assumption
  is not essential and it is done for the clarity of argument.  For the general supermanifold ${\cal M}$ we will have
   to apply the exponential map in order to map the problem to the vector space and keep the covariance.
    The example of this full covariant construction will be given when we discuss the Rozansky-Witten theory
     in section \ref{sec_RW_model}.

\subsection{Gauge fixing}

We continue to discuss the AKSZ model of the previous section. In
general it  may be tricky to pick a Lagrangian subspace ${\cal L}$ such
that the restriction of the action to ${\cal L}$ has a
non-degenerate quadratic term. In our case we expand out the kinetic
term from the action (\ref{AKSZ_action})
\bea S_{kin}=\int\limits_{\Sigma_3}
d^3\xi~\Omega_{AB}\big(-{X}^A_{(1)}\wedge d{X}^B_{(1)}+{X}^A_{(2)}\wedge d{X}^B_{(0)}\big)~,\nn\eea%
where we have assumed that we are in a Darboux coordinate so that
$\Omega$ is a constant. So the kinetic term to be inverted is the
de Rham differential, which has infinitely many zero modes.
It is an intricate game trying to find a set of constraints upon the
component fields such that we are able to invert $d$. However, a
 brutal gauge fixing is possible and works for all model at
the cost of explicit covariance on ${\cal M}$ and this is why we discuss
 the case when ${\cal M}$ is a vector space.

To this end, we introduce a metric on the source manifold
$\Sigma_3$. Using this metric we can use the Hodge decomposition
to break any differential form on $\Sigma_3$ into three parts%
\bea
\omega=\omega^h+\omega^e+\omega^c=\omega^h+d\tau+d^{\dagger}\lambda~,\nn\eea
where $h$, $e$,  $c$ stand for harmonic, exact and co-exact respectively and
$d^{\dagger}$ is the adjoint of $d$. The three parts are
mutually orthogonal under the following non-degenerate pairing%
\bea(\omega_1,\omega_2)\equiv \int\limits_{\Sigma_3}~
\omega_1\wedge*\omega_2~.\nn\eea Since all the component of a
superfield $\BS{X}^A$ are some differential forms on $\Sigma_3$,
we can decompose them likewise \bea
X^A_{(p)}=(X^A_{(p)})^h+(X^A_{(p)})^e+(X^A_{(p)})^c~.\nn\eea The
trouble maker is the exact part, since they are annihilated by $d$
and infinite in number. Our choice for LagSubMfld ${\cal L}$ will
be to simply stay clear of these exact parts. More concretely, we
first decompose the symplectic form (\ref{sympe2929201}) into \bea
\omega&\sim&\int\limits_{\Sigma_3}d^3x \
\Omega_{AB}\big(\delta({X}^A_{(3)})^h\wedge\delta(X^B_{(0)})^h+\delta(X^A_{(3)})^e\wedge\delta(X^B_{(0)})^c\nn\\%
&&+\delta(X^A_{(1)})^h\wedge\delta(X^B_{(2)})^h+\delta(X^A_{(1)})^e\wedge\delta(X^B_{(2)})^c
+\delta(X^A_{(1)})^c\wedge\delta(X^A_{(2)})^e\big)~.\nn\eea%
Moreover  using the integration by parts there are no
ee and cc combinations and the harmonic part is decoupled from the
rest. Since we are on
a rational homology 3-sphere there are no harmonic terms for the 1 and 2 forms, i.e.
 $(X^A_{(1)})^h= (X^A_{(2)})^h =0$. We make the following gauge choice
 $$(X^A_{(1)})^e=0~,~~~ (X^A_{(2)})^e=0~,~~~X^A_{(3)}=0~, $$
  where we put to zero both harmonic and exact parts of $3$-forms.
  For further discussion we adopt the following notations
$(X^A_{(0)})^h = \textsl{x}_0$ which does not appear in the kinetic term and
 thus corresponds to the zero modes. {\em We will not perform the integral over zero modes!}
  We will treat the zero modes as formal parameters and the integral will be performed only
   over co-exact fields
   \bea
    \int\limits_{{\cal L}}   \BS{F}_0 ... \BS{F}_n ~e^{-S_{kin}}   = \int DX^c_{(0)}  DX^c_{(1)} DX^c_{(2)}~~ \BS{F}_0 ... \BS{F}_n~ e^{-S_{kin}}~,
   \eea
 thus the correlator $\langle \BS{F}_0 ... \BS{F}_n \rangle$ is a function of $\textsl{x}_0 \in {\cal M}$.
  The observable
  \bea
&&\BS{F}\equiv  \int\limits_{\Sigma_3} d^3\xi d^3 \theta ~f\left (\BS{X}(\xi, \theta) \right )\nn\\
&=&\int\limits_{\Sigma_3} d^3\xi~ \left (  X_{(3)}^A~\partial_A f (X_{(0)})  + X_{(2)}^A \wedge X_{(1)}^B~\partial_B \partial_A f(X_{(0)})
 + \frac{1}{6} X_{(1)}^A \wedge X_{(1)}^B \wedge X_{(1)}^C ~\partial_C \partial_B \partial_A f(X_{(0)}) \right )\nn
 \eea
 upon the gauge fixing and Taylor expansion becomes
 \beq
  \BS{F}= \sum\limits_{k=0}^\infty  \frac{1}{k!} \int\limits_{\Sigma_3} d^3\xi~ \left (   X_{(2)}^c
  X_{(1)}^c   (X^c_{(0)})^k ~ \partial^{2+k} f(\textsl{x}_0) + \frac{1}{6}   (X_{(1)}^c)^3(X^c_{(0)})^k ~\partial^{3+k}  f(\textsl{x}_0) \right )~,\label{obser3839}
 \eea
 where we suppressed all indices and wedges. Now we have to contract the co-exact fields
  according to the Wick theorem. Using the propagator which is proportional to $\Omega^{-1}$
  the fields $ X_{(1)}^c$ are  constructed to  $X_{(1)}^c$ and the fields $ X_{(2)}^c$ are contracted to
   $ X_{(0)}^c$,  namely
   \bea
     \langle X_{(2)}^A(\xi_1) ~X_{(0)}^B (\xi_2)\rangle& =& (\Omega^{-1})^{AB}~ G^0 (\xi_1, \xi_2)~,\label{VER111}\\
    \langle X_{(1)}^A(\xi_1) ~X_{(1)}^B (\xi_2)\rangle & = &(\Omega^{-1})^{AB}~ G^1 (\xi_1, \xi_2)~, \label{VER222}
   \eea
    where we suppressed superscript $c$ on the fields.
   Since the observable (\ref{obser3839}) is composite in terms of elementary fields, we may have to
    use the point splitting procedure and study the tadpole contributions, if they are there.
     Equivalently we may work  with the superfields and develop the perturbation theory entirely in terms of superfields, thus avoiding
  the components. For this we have to introduce  the adjoint of $D$ which  can be written as
$D^{\dagger}=\nabla^a\partial_{\theta^a}$. The superfield admits the Hodge decomposition with respect to $D$ and $D^\dagger$.
Thus the gauge fixing corresponds to setting to zero the exact part of superfield.

 We leave the explicit formulas for the propogators for the later discussion in section \ref{sec_Min_model}.
  Now we would like to concentrate on the general features of the Feynman rules.

\subsection{Feynman rules}

The current gauge fixing is very
explicit, allowing us to sharpen some features of the perturbation
expansion. The main consequence of this gauge fixing is that
\emph{every diagram will have even number of vertices, all of
which are 3-valent and there are no tadpoles}.

As discussed in the previous subsection the Feynman rules  are defined
 by the $(k+2)$- and $(k+3)$- valent vertices (\ref{obser3839}) and the propogators
  (\ref{VER111}, \ref{VER222}). The first observation is that there will be no $2$-valent vertices
   since
   \bea
    \int\limits_{\Sigma_3} d^3\xi ~  X_{(2)}^c  X_{(1)}^c~\partial^{2} f(\textsl{x}_0)  =0
\eea
 is identically zero\footnote{One may wonder that this argument is too rough. At best we can say that $2$-valent vertex
   $\int d^3\xi ~  X_{(2)}^c  X_{(1)}^c \partial^{2} f(\textsl{x}_0) $ is a surface term and due to the singularities in the propagators
    there  may be non-trivial contributions of $2$-valent vertices.
 However one may perform more careful analysis taking into account the possible singularities of the propagator and arrive at
  the same conclusion that $2$-valent vertices do not contribute.}.
For the next observation it can be useful to think about the Feynman rules in terms of superfields.
 Let us look at the correlator
\bea
 \int\limits_{{\cal L}}   \BS{F}_0 ... \BS{F}_n ~e^{-S_{kin}}  =\int\limits_{\cal L}{\Large\textrm{$\smallint$}}d^6z_0\ f_0(z_0)\cdots
{\Large\textrm{$\smallint$}}d^6z_n\ f_n(z_n)\ e^{-S_{kin}}~,\nn
\eea%
where $f_i(z_i)=f_i(\BS X(z_i))$.  We have a total of ($n+1$)
$\int d^3\theta$. The propogator $\langle \BS{X}(z_1)
\BS{X}(z_2)\rangle$ is
 quadratic in $\theta$'s.  Since we have to saturate all $\theta$-integration there is
  the following relation between the number ${\cal V}$ of vertices and the number ${\cal P}$ of propagators
\bea
 3 {\cal V} = 2 {\cal P}~.\label{PSV999}
\eea
 Equivalently we can make the similar argument within the component form of the perturbation theory.
 The integration over
the configuration space $\Sigma_3 \times ... \times \Sigma_3$ requires a $3(n+1)$-form.
The propagator is a 2-form\footnote{Strictly speaking the correlator is $2$-form on $(\Sigma_3 \times \Sigma_3- \rm{diagonal})$.
 However the singularity of the propogator along the diagonal is not enough to spoil the argument.} on $\Sigma_3 \times \Sigma_3$ since according to
 (\ref{kinetic_comp}) we have propagators between
$X_{(1)},X_{(1)}$ and between $X_{(0)},X_{(2)}$.  Thus to absorb all integration we have to require (\ref{PSV999}).

 The property (\ref{PSV999}) says that in order for the diagram to be non-zero there should be
  $3$-valent vertices on the average. For example, if there is $4$-valent vertex then it should accompanied by
   $2$-valent vertex. However we have argued that $2$-valent vertices vanish identically. Therefore we can conclude
    that only $3$-valent vertices contribute
      \beq
&  \int\limits_{\Sigma_3}& d^3\xi~  X^A_{(0)} X_{(2)}^B\wedge
  X_{(1)}^C~  \partial_C \partial_B \partial_A f(\textsl{x}_0)   ~,\label{obser3839ffkkfdii3}\\
  \frac{1}{6} &\int\limits_{\Sigma_3} & d^3\xi~  X_{(1)}^A \wedge
      X_{(1)}^B \wedge X_{(1)}^C~ \partial_C \partial_B \partial_A f(\textsl{x}_0) ~.\label{obser3839ffkkf}
 \eea
  Since only $3$-valent graphs contribute, we need the even number of $3$-valent vertices to contract all legs.
   Therefore only graphs with the even number of vertices give non-zero contribution.
 Now we have to discuss the tadpoles, the situation when the vertex leg is contracted with another leg from the same vertex.
  The tadpoles contain the following contribution
  \bea
    \partial_A \partial_B \partial_C f(\textsl{x}_0) ~ (\Omega^{-1})^{AB}~,\nn
  \eea
  which is identically zero due to the fact that we contract (graded) symmetric combination $ \partial_A \partial_B$ with
   (graded) antisymmetric $ (\Omega^{-1})^{AB}$.  Thus all tadpoles are automatically zero in the theory\footnote{It is important to stress
    that on $\Sigma_3$ we can do systematically the point-splitting regularization by picking nowhere vanishing vector field.}.

  Thus the correlator $\langle \BS{F}_0 \BS{F}_1 ... \BS{F}_n \rangle$    has the  form $\sum\limits_{\Gamma} b_{\Gamma} I_{\Gamma}$, where
   $\Gamma$'s are all $3$-valent graphs with $(n+1)$ vertices. The number $b_{\Gamma}$ is an integral of the collection
    of propagators $G$ over
    $\Sigma_3 \times ... \times \Sigma_3$ dictated by the graph $\Gamma$. While $I_{\Gamma}$ is collection of
     third derivatives $\partial^3 f_i (\textsl{x}_0)$ contracted by $\Omega^{-1}$ in the way dictated by the $3$-valent
      graph $\Gamma$.  Thus $I_{\Gamma}$ is a function of  zero modes $\textsl{x}_0$.  The explicit example of
       the calculation of the correlator is presented later.

\subsection{The properties of correlators}

In previous subsection we discussed the calculation of the correlators in the perturbative theory corresponding to
 3D AKSZ models. Now we would like to go back to our formal BV arguments about the properties of the correlators, see
  subsections \ref{sub-formcor1} and \ref{sub-formcor2}. We want to understand if those arguments are applicable to
   the perturbative theory, maybe with the possible refinements.

 The perturbative correlator associated with the the collection of functions $f_0, f_1, ... , f_n$ on target ${\cal M}$ is
  defined as follows
\bea c^n( \BB{X}_{f_0}\wedge \BB{X}_{ f_1} \wedge ... \wedge
\BB{X}_{f_{n}})= \langle \BS{F}_0 \BS{F}_1 ...
\BS{F}_n\rangle=\sum_{\Gamma}b_{\Gamma}I_{\Gamma}
(\textsl{x}_0)~,\label{BgammaI3838ma} \eea
 and it depends on zero modes parametrized by ${\cal M}$ itself.  We choose not to integrate over zero modes
  due to the fact that they do not enter the perturbative theory.  Moreover quite often the integration over
   zero modes is either not well-defined or even when it is defined we may miss some interesting structures
    if we perform the integral right away.  From the formal BV arguments we expect that (\ref{BgammaI3838ma})
     is cocycle of Lie algebra of Hamiltonian vector fields on ${\cal M}$.  However now $c^n$ is a function on
      ${\cal M}$ itself. Thus we are dealing with the cochain $c^n$ taking values in the function on ${\cal M}$ and
       one would expect that the differential $\delta$ should be modified. The natural modification looks as follows
\bea
(\delta
c^n)(\BB{X}_{f_0}\wedge\cdots\BB{X}_{f_{n+1}})
&=&\sum_{i<j}(-1)^{l_{ij}} c^n(\BB{X}_{\{f_i,f_j\}}\wedge\cdots\widehat{\BB{X}}_{f_i}
\wedge\cdots\widehat{\BB{X}}_{f_j}\wedge\cdots\BB{X}_{f_{n+1}})\nn\\%
&&-\sum_{i}(-1)^{p_{i}}\big\{f_i(\textsl{x}_0),
c^n(\BB{X}_{f_0}\wedge\cdots\widehat{\BB{X}}_{f_i}
\wedge\cdots\BB{X}_{f_{n+1}})\big\}~,\label{general_diff3838}\eea%
 where we assume that $(n+1)$ is even and the following sign conventions are valid
\bea
l_{ij} &=&f_i+(f_i+3)(f_0+\cdots
f_{i-1}+i3)+(f_j+3)(f_0+\cdots
f_{j-1}+j3)+(f_i+3)(f_j+3)~,\nn \\
p_{i}&=&(f_i+3)(f_0+\cdots f_{i-1}+i3)~. \nn
\eea
 Indeed the formula (\ref{general_diff3838}) can be derived from the first principle with the BV framework.
  We have to treat carefully the contribution of "zero modes" to the  odd Laplacian operator and a
   regularization of the odd Laplacian is required in order to make some of the manipulations
    well-defined. It all can be done and we present the BV derivation of the formula (\ref{general_diff3838}) in the Appendix.
     The final claim is that  the perturbative correlator is cocycle with the values in functions on ${\cal M}$.
     \bea
  (\delta c^n)(\BB{X}_{f_0}\wedge\cdots\BB{X}_{f_{n+1}})= \delta \left (\sum\limits_{\Gamma} b_\Gamma  I_{\Gamma}(\textsl{x}_0) \right )=0 ~,
  \label{djdjd9339312}
     \eea
     where $\delta$ is defined by the formula (\ref{general_diff3838}).

    In order to avoid the discussion of the covariance on ${\cal M}$ we consider the case when ${\cal M} = \mathbb{R}^{2m|k}$
     with even constant symplectic structure $\Omega$.  In this case
  the correlator (\ref{BgammaI3838ma})  is automatically $osp(2m|k, \mathbb{R})$ invariant  since our model has global
   $osp(2m|k, \mathbb{R})$ symmetry.  We believe that the property (\ref{djdjd9339312}) is true for any functions
    $f_i$ on $\mathbb{R}^{2m|k}$ with the constant even symplectic structure (see subsection \ref{sec_Min_model} for some simple
     explicit check).  Now if we fix the point $\textsl{x}_0$ (e.g, choose it to be an origin $\textsl{x}_0=0$) and consider
      the polynomial functions $f_i$ with property $f_i (\textsl{x}_0)= \partial f_i (\textsl{x}_0)=0$, namely members of $\textbf{Ham}^0_{2m|k}$, then the last term
       (\ref{general_diff3838}) disappears and we can regard $c^n$ as a cocycle with the values in ${\mathbb R}$.
   Since the construction of $c^n$ as $\sum b_\Gamma I_{\Gamma}$  is $osp(2m|k, \mathbb{R})$-invariant   we get that
    $c^n$ is the representative of the relative cohomology class
\bea
H^{\bullet}({\mathbf{Ham}}_{2m|k}^0, osp(2m|k, \mathbb{R}), \mathbb{R}) ~.\nn
\eea
   Another observation by Kontsevich  which is natural within the present BV context is that
the following \emph{co}chain in the graph complex%
\bea \sum_{\Gamma}b_{\Gamma}\Gamma^*\nn\eea%
is a \emph{closed}.  The rough proof of this goes as follows.
 We take any graph chain in ${\cal G}_{\bullet}$ and find the
corresponding chain in the CE complex $C_\bullet
(\mathbf{Ham}^0_{2m|k})\equiv
 \wedge^{\bullet}~ \mathbf{Ham}^0_{2m|k} $ according to
section \ref{sec_review}. Evaluating this CE chain in the path
integral results in the number $b_{\Gamma}$, which shows
$b_{\Gamma}$ is a graph cocycle. For a more careful proof of this
statement see reference \cite{Hamilton:2007tg}.
 We can illustrate this argument by the following example.  Recall from
 the equation (\ref{LC_GC_Correspondence})  the correspondence between the graph
complex and the CE complex $C_\bullet (\mathbf{Ham}^0_{2m|k})$.   Suppose we are given the following
3-valent graph with two vertices
\\
\\
\bea \Gamma=\feyn{{f fl flu f}}  \label{simplegraph11}
\eea
\\
\\
To construct a cochain mapping this graph to the number field, one
can first map it to an element of CE complex%
\bea
 \Gamma~\longrightarrow~
\Omega_{A B}\Omega_{CD}\Omega_{KL}~
\BB{X}_{x^{A}x^{C}x^{K}}\wedge\BB{X}_{x^{L}x^{D}x^{B}}~,\nn\eea
and then evaluate this Lie algebra chain in the path integral
according to equation (\ref{requ}). If one defines the two point
correlator as%
\bea
\langle
\BS{X}^{A}(z_1) \BS{X}^{B}(x_2)\rangle \equiv \Omega^{AB}~G(z_1,z_2)~,\nn
\eea
then the result of evaluating the Lie algebra chain is%
\bea
\Omega_{A B}\Omega_{CD}\Omega_{KL}~~
c^1 \left ( \BB{X}_{x^{A}x^{C}x^{K}}\wedge\BB{X}_{x^{L}x^{D}x^{B}} \right )
\sim \int\limits_{T[1]\Sigma_3}d^6z_1d^6z_2~
G(z_1,z_2)^3\equiv b_{\Gamma}~.\nn\eea%
Since the mapping between the graph complex and the CE chain
complex is an isomorphism, and moreover the path integral is a
cocycle in the CE cochain complex, we conclude that
$\sum_{\Gamma}b_{\Gamma}\Gamma^*$ is a cocycle.

\subsection{Partition function}

 In this subsection we study the perturbative partition function for
  the action (\ref{AKSZ_action})  with the interaction term.  We want
   to apply the general ideas about the characteristic classes of flat bundles
   reviewed in subsection \ref{sub-charclass} to the perturbative partition function
    of 3D AKSZ models.

   Using the gauge fixing and Feynman rules  from the previous subsections the partition
    function has the following expansion
  \bea
   Z(\textsl{x}_0) = \sum\limits_{n=0}^\infty \frac{(-1)^n}{n!} c^{n-1} (\BB{X}_\Theta \wedge ... \wedge \BB{X}_\Theta) ~,\label{partfunc2727}
  \eea
   where $c^n$ is cocycle evaluated at $\Theta$
\bea c^n (\BB{X}_\Theta \wedge ... \wedge \BB{X}_\Theta) \equiv
\int\limits_{\cal
L}{\Large\textrm{$\smallint$}}d^6z_0{\BS\Theta}(z_0)\cdots
{\Large\textrm{$\smallint$}}d^6z_n ~\BS{\Theta}(z_n)~e^{-S_{kin}}
= \sum\limits_{\Gamma} b_\Gamma c_{\Gamma}
(\textsl{x}_0)~,\label{disconnected} \eea
 with $c_{\Gamma}(\textsl{x}_0)$ is constructed by contracting $\partial^3 \Theta (\textsl{x}_0)$ with $\Omega^{-1}$ according
  to the graph $\Gamma$.  We stress that we do not perform the integral over zero modes which are parametrized by the target ${\cal M}$
   and thus our partition function is function of $\textsl{x}_0$. However in the Chern-Simons theory $\Theta$ is cubic in fields  and
thus $\partial^3 \Theta (\textsl{x}_0)$ are constants and we end up with the constant partition function, in this particular example.

Since the function $\Theta$ on ${\cal M}$ satisfies $\{ \Theta, \Theta\}=0$ there is homological vector field
 $Q^A (\textsl{x}_0) = \Omega^{AB} \partial_B \Theta$, which squares to zero (i.e., the Lie bracket $Q^A \partial_A Q^B =0$).
  Thus there is natural differential acting on the functions on ${\cal M}$ and we can define a cohomology group $H_Q({\cal M})$.
   We would like to argue that $Q \cdot c^n=0$ and as a result $Q \cdot Z(\textsl{x}_0)=0$.  Therefore the partition function $Z(\textsl{x}_0)$
    can be understood as some sort of $Q$-characteristic class, i.e. the element of $H_Q({\cal M})$.

 Let us start from very elementary proof of this fact. As we argued before $c^n(\textsl{x}_0)$ is cocycle with respect to the differential
  $\delta$ defined by the formula (\ref{general_diff3838}).  Thus we have the following chain of relations
  \bea
   Q \cdot c^n = \{\Theta(\textsl{x}_0), c^n \} \sim  c^n (\BB{X}_{\{\Theta, \Theta\}} \wedge \BB{X}_\Theta \wedge ... \wedge \BB{X}_\Theta) =0~\label{Q_class},
  \eea
   where we used $\delta c^n=0$ and $\{\Theta, \Theta \}=0$.  Although this derivation is correct and simple,
    it misses some important geometrical aspects.  Now we will resolve to more elaborate argument, but with clear geometrical
     meaning.

For simplicity, we take the target space ${\cal M}$  to be a
vector space and identify its tangent space with the
manifold itself $T{\cal M} = {\cal M} \times {\cal M}$,  both are equipped with a symplectic structure
and a bracket. Let us try to compute the partition function $Z(\textsl{x}_0)$ of the
AKSZ model. The standard method is to split any field into the
classical part and a fluctuation part $X=\textsl{x}_0+\xi$.
$\textsl{x}_0$ is treated as the background while the fluctuation
$\xi$ is taken to parameterize the fiber of the tangent bundle of
${\cal M}$ at $\textsl{x}_0$. For a general curved manifold the simple
splitting $\textsl{x}_0+\xi$ does not make sense, one has to use
the exponential map to identify a neighborhood of the tangent
bundle with the neighborhood of $\textsl{x}_0$; this is what we do
in section \ref{sec_RW_model}.

From the master equation $\{\Theta,\Theta\}=0$, we want to derive
some sort of Cartan-Maurer equation for the bundle $T{\cal M}$. We Taylor
expand $\{\Theta,\Theta\}=0$ around $\textsl{x}_0$ in powers of
$\xi$, and get a series of equations%
\bea
0&=&2(\partial_{C}\partial_A\Theta)(\Omega^{-1})^{AB}\partial_{B}\Theta~,\nn\\%
0&=&(\partial_A\Theta)(\Omega^{-1})^{AB}\partial_{C_1}\partial_{C_2}\partial_{B}\Theta+
(\partial_{C_1}\partial_A\Theta)(\Omega^{-1})^{AB}\partial_{C_2}\partial_{B}\Theta~,\nn\\%
&\cdots&\nn\\%
0&=&\frac{2}{n!}(\partial_A\Theta)(\Omega^{-1})^{AB}\partial_{C_1}\cdots\partial_{C_n}\partial_{B}\Theta\nn\\%
&&+\frac{1}{n!}
\sum_{p=1}^{n-1} \binom{n}{p} (\partial_{C_1}\cdots\partial_{C_p}\partial_A\Theta)(\Omega^{-1})^{AB}
\partial_{C_{p+1}}\cdots\partial_{C_n}\partial_{B}\Theta~.\label{series_Eqn}\eea
If we define $\Theta'$ as
\bea \Theta'=\sum_{n=2}^{\infty}\frac{1}{n!}(\partial_{C_1}\cdots
\partial_{C_n}\Theta(\textsl{x}_0))\xi^{C_1}\cdots\xi^{C_n}~,\nn\eea
the series of equation except the first one can be packaged into
the following compact form%
\bea
(\partial_A\Theta(\textsl{x}_0))(\Omega^{-1})^{AB}\frac{\partial}{\partial
\textsl{x}^B_0}\Theta'(\textsl{x}_0)+\frac{1}{2}\{\Theta',\Theta'\}_{\xi}=0~,\nn\eea%
where the bracket is written as $\{,\}_{\xi}$ to stress that it is
the bracket of the fiber of $T{\cal M}$. This is then our favorite
Cartan-Maurer equation%
\bea
Q^B(\textsl{x}_0)\frac{\partial}{\partial
\textsl{x}^B_0}\Theta'(\textsl{x}_0)+\frac{1}{2}\{\Theta',\Theta'\}_{\xi}=0~.\label{gdhdj39skw}\eea
 Now we regard $\Theta'(\textsl{x}_0, \xi)$ as a function on $T{\cal M} = {\cal M} \times {\cal M}$.
  Since the zero term in expansion $\Theta (\textsl{x}_0)$ will not be joined by the
propagators, the connected diagrams in $c^n$  from (\ref{partfunc2727})  will be given by
\bea
 \int\limits_{\cal L}
{\Large\textrm{$\smallint$}}d^6z_0\BS{\Theta}'(z_0)\cdots{\Large\textrm{$\smallint$}}
d^6z_n\BS{\Theta}'(z_n)~e^{-S_{kin}}~,\label{connected123}
\eeq
   which can be regarded as cocyle of Lie algebra formal Hamiltonian vector fields
    along the fiber $T{\cal M}$ (i.e., $\xi$-direction).
    By construction $\Theta' (\textsl{x}_0, 0)=0$ and $\partial_\xi \Theta' (\textsl{x}_0, 0)=0$ and thus
     the expression (\ref{connected123}) is annihilated by $\delta$ as defined in (\ref{general_diff3838}),
     but without the last term.  From (\ref{gdhdj39skw}) $\Theta'$ can be thought of as flat connection on bundle
      ${\cal M}\times{\cal M}$ with the Lie algebra being a Lie algebra of formal vector fields  ${\mathbf{Ham}}_{2m|k}^0$
       and the base differential being $Q$. The perturbative calculation can be understood as plugging the flat connection
        $\Theta'$ into a cocycle $c^n$ and obtaining the characteristic class  in $H_Q({\cal M})$ of the flat bundle ${\cal M} \times {\cal M}$.
 Thus according to the general discussion around the equation (\ref{general_CM}),
the correlator $\langle\int \Theta'\cdots \int \Theta'\rangle$
will be annihilated by $Q^A(\textsl{x}_0)$. But it may be helpful to
understand this in a concrete context. If we adopt the particular
gauge fixing given in section \ref{sec_gauge_fix}, then $\textsl{x}_0$
is naturally taken to be the harmonic 0-form part of the fields
$X^h_{(0)}$, $\xi$ is the non-harmonic part $\xi\sim
\BS{X}^e+\BS{X}^c$. Recall from the previous discussion, the perturbation expansion picks up only
the part of $\Theta'$ that is cubic in $\xi$, and the correlator
$\langle\int \Theta'\cdots \int \Theta'\rangle$ is given by
tri-valent graphs. Since the higher powers of $\Theta'$ does not
enter the computation, we should be able to understand the
invariance of the correlator under $Q$ in a direct way. Indeed,
using the third equation of (\ref{series_Eqn}), we have%
\bea Q(\partial^3\Theta(\textsl{x}_0)\xi^3)
=-\frac{1}{6}\{\partial^2\Theta\xi^2,\partial^3\Theta
\xi^3\}_{\xi},\nn\eea%
and recall that the quadratic Hamiltonian functions generate the
$osp$ rotations, so $Q$ will act on a correlator
$\langle{\Large\textrm{$\smallint$}}
\partial^3\Theta\xi^3\cdots{\Large\textrm{$\smallint$}}\partial^3\Theta\xi^3\rangle$
as a rotation in the $\xi$ space. But the correlator clearly has
an invariance under such rotations.
  The homomorphism induced by (\ref{gdhdj39skw}) gives us
a mapping%
\bea H^{\bullet}({\mathbf{Ham}}_{2m|k}^0, osp(2m|k, \mathbb{R}),
\mathbb{R}) ~\longrightarrow~H_Q({\cal M})~,\nn \eea which offers
a better explaination for (\ref{Q_class}). We have already
explained the modding out of $osp(2m|k, \mathbb{R})$. Taking the
Lie algebra to be the formal Hamiltonian vector fields has two
advantages, first it has more stable cohomology classes than, say,
the Lie algebra of formal vector fields preserving zero (in which
case, there is only three infinite series of cocycles, see
\cite{lyakhovich-2009,Fuks}). Larger number of cocylces means more
characteristic classes after
 plugging the flat connection into cocycles.  Secondly
any foliation with a symplectic structure transverse to the leaf naturally gives rise to a
 flat
connection taking values in the Lie algebra of Hamiltonian vector fields \cite{FuksI}.
 Indeed any symplectic graded(super)manifold ${\cal M}$ with the
 nilpotent Hamiltonian will give rise to the flat connection on $T{\cal M}$ upon using
  the exponential map.

To summarize, since the partition function of the AKSZ model will
consist of series of (\ref{disconnected}), each of which is
invariant under homological vector field $Q$, the AKSZ model
calculates the characteristic class associated to $Q$.  If the integration over
 zero modes is well-defined then one can calculate the characteristic number
  for the corresponding class.
 This a possible way of producing characteristic classes associated
with a $Q$-structure using path integral for 3D AKSZ models and this agrees with the
prescription given in \cite{lyakhovich-2009} where the authors
showed that basically all that matters are the 3-valent graphs.

Finally let us make comments about the partition function. There
is a dual construction of graph cycles as follows. Suppose we have
an \emph{odd} Hamiltonian function $\Theta$ satisfying
$\{\Theta,\Theta\}=0$, and $\Theta$ is at least quadratic in its
Taylor expansion. We take the cubic or higher Taylor coefficients
as vertices, then we follow the graph and connect the vertices
together using $\Omega^{-1}$, take the graph as in
(\ref{simplegraph11}) (denoting
$\partial_{A}\cdots\partial_{B}\Theta|_{x=0}$ as
$\Theta_{A\cdots B}$)%
\bea \Gamma~\stackrel{\Theta}{\longrightarrow}~
\Omega_{AB}\Omega_{CD}\Omega_{KL}\Theta_{ACK}\Theta_{BDL} \equiv
c_{\Gamma}~,\nn \eea and the chain $\sum_{\Gamma}c_{\Gamma}\Gamma$
is a graph cycle. Now we have seen two dual constructions of the
graph (co)cycle, when we compute the partition function of a
general AKSZ model with the action (\ref{AKSZ_action}), each term
in the
perturbation expansion is of the form%
\bea Z_{AKSZ}=\sum\limits_{n=0}^\infty(-1)^n\frac{1}{n!}\int\limits_{{\cal
L}}{\Large\textrm{$\smallint$}}
d^6z_1\BS{\Theta}(z_1)\cdots{\Large\textrm{$\smallint$}}
d^6z_1\BS{\Theta}(z_n)e^{-S}~,\nn\eea%
complicated as the integral might be, it can always be written as
\bea
\langle\sum_{\Gamma}b_{\Gamma}\Gamma^*,\sum_{\Gamma}c_{\Gamma}\Gamma\rangle=
\sum_{\Gamma}b_{\Gamma}c_{\Gamma}~,\nn\eea%
which realizes the pairing of the two dual constructions.

\section{Example 1: $Q$-Equivariant Bundle}\label{sec_Flat_Bundle}

In this subsection, we give a slightly more concrete example of
the general discussion above, preparing the way for the
Rozansky-Witten model. We consider an example of  Q-equivariant
vector bundle ${\cal E}\rightarrow {\cal M}$ (see the discussion around the equation (\ref{Qequiv}))
 and we set up the corresponding 3D AKSZ model which provides the
realization of the  characteristic classes  described  in
section \ref{sec_review}.

We assume that ${\cal M} = T[1]M$ with $M$ being usual smooth
manifold and the fiber of ${\cal E}$ has a symplectic structure
$\Omega_{AB}$, and we assign to it degree 2. The Lie algebra of
the structure group of the bundle is the Lie algebra of formal
Hamiltonian vector fields. Suppose for definiteness, the
$Q$-structure of the base is the de Rham differential and it is
lifted to a $Q$-structure $\tilde Q$ acting on the total space
${\cal E}$. This action preserves $\Omega_{AB}$ so $\tilde Q$ can
be
written as%
\bea \tilde Q=v^{\mu}\frac{\partial}{\partial
X^{\mu}}+v^{\mu}\{A_{\mu }(X,e),\cdot\}_{\Omega}~,\nn\eea%
for some function $A_{\mu}$ of the total space ${\cal E}$. Of
course, we need $\partial_{[\mu}A_{\nu]}+\{A_{\mu},A_{\nu}\}=0$ to
ensure $\tilde Q^2=0$

To set up an AKSZ model we need a Hamiltonian lift of $\tilde Q$.
We do this within the minimal symplectic realization of  ${\cal E}$. Denote
the local coordinate of this symplectic space to be
$X^{\mu},P_{\mu},v^{\mu},q_{\mu},e^A$, where $e^A$ is the fiber
coordinate of ${\cal E}$ and $v^{\mu}$ is that of $T[1]M$ etc. The degree
assignment is 0 2 1 1 0 respectively. This big space has the
symplectic structure%
\bea \omega=\delta P_{\mu}\wedge \delta X^{\mu}+\delta q_{\mu}
\wedge\delta v^{\mu }+\frac{1}{2}\Omega_{AB}~\delta e^A\wedge\delta
e^B~,\nn\eea
 with $\deg \Omega =2$.
Then we can lift $\tilde Q$
into a Hamiltonian function%
\bea \Theta=P_{\mu}v^{\mu}+v^{\mu}A_{\mu}~.\nn\eea

Using $\Theta$ as the interaction term, we have the standard 3D AKSZ
action%
\bea S=\int\limits_{T[1]\Sigma_3}d^6z\
\BS{P}_{\mu}D\BS{X}^{\mu}+\BS{q}_{\mu}D\BS{v}^{\mu}+\frac{1}{2}\BS{e}^A\Omega_{AB}
D\BS{e}^B+\BS{P}_{\mu}\BS{v}^{\mu}+\BS{v}^{\mu}A_{\mu}(\BS X,\BS{e})~. \label{act_flat_bunle}\eea%
We adopt the same gauge fixing by breaking every field into the
harmonic and non-harmonic parts and setting to zero all the exact
fields. Since $\BS{q}^c$ only appears in the kinetic term
$\BS{q}^cD\BS{v}^c$, we can integrate it out enforcing
$D\BS{v}^c=0$ and hence
$\BS{v}^c=0$, i.e. $\BS{v}$ is harmonic. We are then left with%
\bea S&=&\int\limits_{T[1]\Sigma_3} d^6z\
(\BS{P}_{\mu})^cD(\BS{X}^{\mu})^c+\frac{1}{2}(\BS{e}^A)^c\Omega_{AB}
D(\BS{e}^B)^c\nn\\%
&&\hspace{2cm}+(\BS{v}^{\mu})^h(\BS{P}_{\mu})^h+(\BS{v}^{\mu})^hA_{\mu}(\BS{X}^h+\BS{X}^c,\BS{e}^c+\BS{e}^h)~.\nn\eea
We integrate out $\BS{P}^c$ enforcing $\BS{X}^c=0$:%
\bea S=\int\limits_{T[1]\Sigma_3}d^6z\ \frac{1}{2}\BS{e}^c\Omega
D\BS{e}^c+\BS{v}^h\BS{P}^h+(\BS{v}^{\mu})^h\BS{A}_{\mu}(\BS{X}^h,\BS{e}^c+\BS{e}^h)~.\nn\eea
Remember we are regarding the harmonic fields as parameters rather
than dynamical variables, we can for example put $e^h,P^h$ to
zero, and reduce the action down to its minimal ingredients%
\bea S=\int\limits_{T[1]\Sigma_3}d^6z\ \frac{1}{2}\BS{e}^c\Omega
D\BS{e}^c+(\BS{v}^{\mu})^hA_{\mu}(\BS X^h,\BS{e}^c)~.\nn\eea%
It is then clear from the earlier discussion that the perturbation
expansion will have the Taylor coefficients of
$(\BS{v}^{\mu})^hA_{\mu}(\BS{X}^h,\BS{e}^c)$ as interaction
vertices and the propagators will contract $e^c$'s together. The
result of the path integral is a function of $X^h_{(0)}$ and
$v^h_{(0)}$ only, which is just a differential form on $M$. This
form is closed by construction, and is the secondary
characteristic class.

\section{Example 2: Odd Chern-Simons theory}\label{sec_Min_model}%

Here we present a toy model on a vector space that reproduces the
same weight function $b_{\Gamma}$ as in Chern-Simons and
Rozansky-Witten theory. This model also provides us with cocycles
in Lie algebra cohomology, setting the stage for
section \ref{sec_RW_model}.

\subsection{Chern-Simons theory}

In this subsection we briefly remind some well-known perturbative aspects of the Chern-Simons theory.
 The 3D Chern-Simons theory is defined for any metric Lie algebra ${\mathbf g}$ by the following classical action
\bea S_{CS}=\frac{k}{4\pi}\int\limits_{\Sigma_3}d^3\xi~
\big[ \eta_{\alpha\beta}~ A^\alpha_{(1)}\wedge dA^\beta_{(1)}+\frac{2}{3}f_{\alpha\beta\gamma}~
A^\alpha_{(1)}\wedge A^\beta_{(1)}\wedge A^\gamma_{(1)}\big]~,\label{CSnormal129}\eea%
where $A_{(1)}$ is a connection $1$-form on $\Sigma_3$, $\eta$ is
the metric and $f$ is the structure constant on ${\mathbf g}$.
  The Chern-Simons theory can be embedded into BV-framework through the AKSZ action
\bea
 \int\limits_{T[1]\Sigma_3}d^6z~
\big[\eta_{\alpha\beta}~ \BS{A}^\alpha D\BS{A}^\beta +\frac{2}{3} f_{\alpha\beta\gamma} ~\BS{AAA}\big]~,\label{CSactionBV123}\eea%
where $\BS A$ is a degree 1 superfield valued in a Lie algebra understood as
$$ T[1]\Sigma_3~\longrightarrow~ {\mathbf g}[1]$$
 and the odd symplectic structure is
 \bea
    \int\limits_{T[1]\Sigma_3}d^6z~ \eta_{\alpha\beta}~ \delta \BS{A}^\alpha \wedge \delta \BS{A}^\beta~. \nn
 \eea
  The naive gauge fixing of (\ref{CSactionBV123}) with $A_{(3)} = A_{(2)}=0$ leads to the action (\ref{CSnormal129})
   which is not suited for the perturbative theory. We have to resolve to the gauge fixing we have discussed by setting
    the exact parts of the fields to zero. Namely we get
  \bea
S_{CS}=\int\limits_{\Sigma_3}d^3\xi~\Tr\big[A_{(1)}^c\wedge
dA_{(1)}^c+A_{(2)}^c\wedge
dA_{(0)}^c +\frac{2}{3}A_{(1)}^cA_{(1)}^cA_{(1)}^c+A_{(2)}^c[A_{(1)}^c,A_{(0)}^c]\big]~,\label{CSgaugefixk33}\eea
 where for the sake of clarity we suppressed $\eta$ and $f$.
One can recognize in this action $A_{(1)}$ as the connection
1-form, $A_{(0)}$ as the ghost $c$ and $A_{(2)}$ as the anti-ghost
$d^{\dagger}\bar c$. While the Lagrange multiplier appearing in
 the standard gauge fixing in \cite{Axelrod:1991vq} has been integrated out here forcing
every field to be co-exact.  Therefore our gauge fixing is equivalent to the gauge fixing used in
\cite{Axelrod:1991vq} for the  Chern-Simons action (expanded around a
trivial connection).

If we look at the correlators in the perturbation theory then according to the BV-argument we have to
 get a cocycle of Lie algebra of formal Hamiltonian vector fields on ${\mathbf g}[1]$ with the even symplectic
  structure given by metric $\eta$. The functions on ${\mathbf g}[1]$ are $\wedge^\bullet {\mathbf g}^*$ and
   thus the perturbation theory gives us map
   $$c^n~:~\wedge^\bullet {\mathbf g}^* \otimes \wedge^\bullet {\mathbf g}^* \otimes ... \otimes \wedge^\bullet {\mathbf g}^*~\longrightarrow~ {\mathbb R}~,$$
    which is cocycle with respect to differential $\delta$ defined
    previously. But of course calculating the Lie algebra cocycles is
    hardly the principle use of Chern-Simons theory.

\subsection{Odd Chern-Simons theory}

We take a $2m$ dimensional vector space $M=\mathbb{R}^{2m}$
equipped with the standard symplectic structure $\Omega$. The BV
model will
be $T[1]\Sigma_3\rightarrow M$. The action is the free action%
\bea S=\frac{1}{2}\int\limits_{T[1]\Sigma_3} d^6z~
\BS{X}^\mu\Omega_{\mu\nu} D\BS{X}^\nu~,\nn\eea%
 where we assign formally the symplectic form grading 2 to match
the degree as is done in \cite{AKSZ_RW}.
The model is interesting because if we perform a naive gauge
fixing by setting the 2- and 3-form components of the superfield $\BS
X^\mu$ to zero, we get a component action
\bea
S=\frac{1}{2}\int\limits_{\Sigma_3} d^3\xi~ \Omega_{\mu\nu}
\big(-{X}^\mu_{(1)}\wedge d{X}^\nu_{(1)}\big)~,\nn\eea%
which can be compared  to the free part of Chern-Simons action (\ref{CSnormal129}).
  The only difference between this model and the Chern-Simons is
that the symmetric metric  $\eta_{\alpha\beta}$ is replaced with the
anti-symmetric symplectic form $\Omega_{\mu\nu}$ while the even 1-form
$A_{(1)}^\alpha$ is replaced with the odd 1-forms $X^\mu_{(1)}$. The
similarity does not stop here, as we go on to look at their
perturbation  expansion. We will refer to this new theory as  odd Chern-Simons theory.

We use the gauge fixing of the previous section. The resulting
action is%
\bea \int\limits_{\Sigma_3}d^3\xi~
\Omega_{\mu\nu}\big((X^\mu_{(2)})^c\wedge
d(X^\nu_{(0)})^c-(X^\mu_{(1)})^c\wedge
d(X^\nu_{(1)})^c\big)~,\nn\eea which again  can be compared to
free part of  the gauge fixed Chern-Simons action
(\ref{CSgaugefixk33}).

 We want to discuss the perturbative theory for odd Chern-Simons model.
From the above consideration, it is not surprising that, as far as the
weight function $b_{\Gamma}$ is concerned, these two models (odd and even Chern-Simons) are
equivalent. And what is more, although the Rozansky-Witten model,
being an AKSZ model was gauge fixed slightly differently, also
produces the same weight function. Next we look at a two point
function show the total agreement between the odd CS model, the CS
model and the Rozansky-Witten model.

The Green's function may be worked out in a conventional manner,
we first insert sources for the fields and compute the partition
function%
\bea Z[J]=\int{\cal D}\BS{X}\exp\Big\{\Large{\textrm{$\smallint$}}
d^3\xi~ X_{(2)}\Omega
dX_{(0)}-X_{(1)}\Omega dX_{(1)}+J_{(1)}X_{(2)}+J_{(2)}X_{(1)}+J_{(3)}X_{(0)}\Big\}~.\nn\eea%
Complete the square for the action (watch out $J_{(2)}$ is odd
$J_{(1)},J_{(3)}$ are even)%
\bea
S[J]&=&\int d^3\xi~\big((X_{(2)}-\frac{1}{d}J_{(3)}\Omega^{-1})
\Omega(dX_{(0)}+\Omega^{-1}J_{(1)})-J_{(1)}\Omega^{-1}\frac{1}{d}J_{(3)}\nn\\%
&&-(X_{(1)}+\frac{1}{2d}J_{(2)}\Omega^{-1})\Omega(dX_{(1)}-\Omega^{-1}\frac{1}{2}J_{(2)})-\frac{1}{4}J_{(2)}
\Omega^{-1}\frac{1}{d}J_{(2)}\big)~.\nn\eea%
So the partition function is%
\bea Z[J]&=&\frac{1}{\Delta}\exp\int
d^3\xi~\big(-J_{(1)}\Omega^{-1}\frac{1}{d}J_{(3)}-\frac{1}{4}J_{(2)}
\Omega^{-1}\frac{1}{d}J_{(2)}\big)~,\nn\eea%
where we used $\Delta$ to denote the 1-loop determinant factor.
Note that the absolute value of it is the Ray-Singer torsion and
it is independent of the metric on $\Sigma_3$. The phase of
$\Delta$ is much more delicate, according to
 \cite{Witten:1988hf}, a gravitational Chern-Simons term must
be added to the phase factor of $\Delta$ to restore the metric
independence.

The Green's functions for $J_{(2)},J_{(3)}$ are
\bea&&\frac{1}{d}J_3(u)=*d_u\int dv\;G(u,v)*J_3(v)g^{1/2}~,\nn\\%
&&\frac{1}{d}J_2(u)=\int dv\;H^{\ bc}_a(u,v)J_{bc}(v)g^{1/2}~,\nn\eea%
where $G(u,v)$ is just the scalar Green's function satisfying
$\nabla^2_uG(u,v)=\delta(u,v)/\sqrt{g}$. We do not have the
Green's function for $J_{(2)}$ over arbitrary $\Sigma_3$, so we
write it as $H_a^{\ bc}(u,v)$ in general. But in flat $\Sigma_3$,
the Green's function is $\int
G(u,v)d^{\dagger}J_{(2)}$.

The correlators are obtained by varying $Z[J]$ with respect to  the source%
\bea (\Omega^{-1}) G^0_{ab}(u,v)&=&\langle
X_{ab}(u),X (v)\rangle=\epsilon_{abd}\frac{\partial}{\partial
J_d(u)}\frac{\partial}{\partial J_{(3)}(v)}Z[J]\nn\\%
&=&-g^{1/2}(u)\partial_{u^c}G(u,v)\epsilon^c_{\ ab}~,\nn\\%
(\Omega^{-1}) G^1_{ab}(u,v)&=&\langle
X_a(u),X_b(v)\rangle=\frac{1}{4}\epsilon_{acd}\epsilon_{bef}\frac{\partial}{\partial
J_{cd}(u)}\frac{\partial}{\partial J_{ef}(v)}Z[J]\nn\\%
&=&H_a^{\ cd}(u,v)g^{1/2}\epsilon_{bcd}~,\eea
 where we suppressed the target space indices.
We can assemble the Green's function into the superfield form%
\bea\langle\BS{X}(u,\theta),\BS{X}(v,\eta)\rangle=\frac{1}{2}\theta^b\theta^a
G_{ab}^0(u,v)-\theta^a\eta^bG_{ab}^1(u,v)+\frac{1}{2}\eta^b\eta^a
G_{ab}^0(v,u)\nn\eea%
Note in the limit $\Sigma_3$ is flat, the $G_{ab}^1$ is given by
$\epsilon_{ab}^{\ \ c}\partial_cG(u,v)$, so the super Green's
function becomes%
\bea\langle\BS{X}(u,\theta),\BS{X}(v,\eta)\rangle=-\frac{1}{2}\theta^b\theta^a
\epsilon_{ab}^{\ \
c}\partial_cG(u,v)-\theta^a\eta^b\epsilon_{ab}^{\ \
c}\partial_cG(u,v)-\frac{1}{2}\eta^b\eta^a \epsilon_{ab}^{\ \
c}\partial_cG(u,v)\sim (\theta-\eta)^2\nn\eea%
hence there can not be two propagators between two vertices in the
flat limit. This is important for this accounts for the improved
short distance behavior. Since one propagator blows up with the
square inverse of distance between two insertions, so there will
be a naive UV divergence when three propagators connect two
insertions. Yet, as two insertions become coincident, the flat
space propagator dominates and we just saw the mitigation of such
UV behavior. In fact ref.\cite{Axelrod:1991vq} proved the
finiteness of the perturbation expansion.

Now we try to compute the two point function
 \bea
 c^1 (\BB{X}_{f_1} \wedge \BB{X}_{f_2})=\int\limits_{\cal L}
\Large{\textrm{$\smallint$}} d^6z
f_1\;\Large{\textrm{$\smallint$}} d^6z f_2\;e^{-S}~.\nn\eea%
There is only one 3-valent diagram
\\
\bea \feyn{\vertexlabel^{f_1}\  {f fl flu f}\ \vertexlabel^{\ \;f_2}}\nn
\\ \nn
\eea%
in order to have enough $\theta$'s to satisfy
the Grassmann integral. The correlator is formally written as%
\bea
 c^1 (\BB{X}_{f_1} \wedge \BB{X}_{f_2}) =b_{\Gamma}\cdot
(\Omega^{-1})^{\mu_1\nu_1}(\Omega^{-1})^{\mu_2\nu_2}(\Omega^{-1})^{\mu_3\nu_3}
(\partial_{\mu_1}\partial_{\mu_2}\partial_{\mu_3} f_1)(\partial_{\nu_1}\partial_{\nu_2}\partial_{\nu_3}f_2)~,\label{corr3738}
\eea
where $b_{\Gamma}$ is the weight function\footnote{Our convention
is that $e^{abc},e_{abc}=\epsilon_{abc}$ is the Levi-Civita
symbol, while $\epsilon^{abc}=1/ge^{abc}$}%
\bea b_{\Gamma}&=&\int d^3ud^3\theta\int
d^3vd^3\eta~\big(\frac{1}{2}\theta^b\theta^a
G_{ab}^0(u,v)-\theta^a\eta^bG_{ab}^1(u,v)+\frac{1}{2}\eta^b\eta^a G_{ab}^0(v,u)\big)^3\nn\\%
&=&-\int d^3ud^3\theta
d^3vd^3\eta~\big(\frac{6}{4}(\theta^a\theta^bG^0(u,v))(\theta^a\eta^bG_{ab}^1(u,v))(\theta^a\theta^b
G^0(u,v))+(\theta^a\eta^bG_{ab}^1(u,v))^3\big)\nn\\%
&=&-\int d^3u d^3v~\big(\frac{6}{4}G_{ab}^0(u,v)G_{cd}^1(u,v)
G_{ef}^0(v,u)+G_{ab}^1(u,v)G_{cd}^1(u,v)G^1_{ef}(u,v)\big)
e^{abc}e^{def}\nn\\%
&=&-\int d^3u d^3v~\big(6g^{1/2}(u)\partial^c_uG(u,v)G_{cd}^1(u,v)
\partial_v^dG(v,u)g^{1/2}(v)\nn\\%
&&+G_{ab}^1(u,v)G_{cd}^1(u,v)G^1_{ef}(u,v)e^{abc}e^{def}\big)~.\nn\eea
We observe that this is the same weight function appearing in the
CS perturbation theory \cite{Axelrod:1991vq} and in the
Rozansky-Witten model \cite{Rozansky:1996bq}. And we expect the
agreement\footnote{Indeed we expect that the same agreement for  $b_\Gamma$ for the generic theory
with the target ${\mathbb R}^{2m|k}$.} to continue even for larger diagrams.  The metric dependence of $b_\Gamma$ can
 be addressed in the same way as for the Chern-Simons theory and thus we leave this issue aside.

For this simple correlator (\ref{corr3738}) we may check explicitly that it is cocycle
\bea
 (\delta c^1) (\BB{X}_{f_1} \wedge \BB{X}_{f_2} \wedge \BB{X}_{f_3}) = c^1 ( \BB{X}_{\{f_1, f_2\}} \wedge \BB{X}_{f_3} )-
 c^1 ( \BB{X}_{\{f_1, f_3\}} \wedge \BB{X}_{f_2} ) + c^1 ( \BB{X}_{\{f_2, f_3\}} \wedge \BB{X}_{f_1} ) \nn \\
 - \{ f_1, c^1 (\BB{X}_{f_2} \wedge \BB{X}_{f_3} ) \} + \{ f_2, c^1 (\BB{X}_{f_1} \wedge \BB{X}_{f_3} ) \}
 +  \{ f_3, c^1 (\BB{X}_{f_1} \wedge \BB{X}_{f_2} )\} =0~, \nn
\eea
 where magically all derivatives cancel out for any functions $f_1, f_2, f_3$.

Some comments must be made regarding the subtlety of the (source)
metric independence. The original model is written down without
any metric and hence is classically invariant under any
orientation preserving diffeomorphism of $\Sigma_3$. The metric
only comes in through the gauge fixing we use, and so the
correlator should not depend on the gauge choice by the discussion
of section \ref{sec_BV} (known as the the Ward-identity in gauge
theory). However, caution is needed when making such assertions,
for the Ward-identity maybe spoiled by the quantum correction and
we already saw that the 1-loop determinant has an anomalous
dependence on the metric. Even when a correlator is finite
accidentally, like the two point function above, one cannot
conclude based on finiteness its gauge invariance, indeed
\cite{Axelrod:1991vq} showed that this two loop diagram
suffers a similar anomaly as the determinant factor. Another
somewhat remote example for this is the 1-loop light by light
scattering in QED. The diagrams when computed in 4D are finite
accidentally, yet the result is \emph{not} gauge invariant. The
usual wisdom for the gauge theory is that the question hangs upon
whether one possesses a gauge invariant regulator, if yes, then
one can subtract the divergence in a manner preserving the
symmetry in question and the symmetry is anomaly free. For the CS
model, the common practice is that when one integrates over the
configuration space, which is copies of $T[1]\Sigma_3$, one
carefully subtracts the diagonals because these correspond to the
singular configuration where two insertions are coincident. As a
result one is led to consider certain compactification of the
configuration space, but this is out of the scope of this paper.

\section{Example 3: Reinterpreting Rozansky-Witten Model}\label{sec_RW_model}%

The Rozansky-Witten (RW) model was introduced in
 \cite{Rozansky:1996bq} and it gives rise to the Rozansky-Witten invariants,
  see \cite{sawon} for the nice review of these invariants.
 The authors of \cite{Rozansky:1996bq}
constructed the model by writing down a set of BRST rules
associated to a hyperK\"ahler manifold. They also pointed out the
similarity between their model and the Chern-Simons model and went
as far as calling it the odd Chern-Simons model. Yet one important
difference between the two is that the perturbation expansion of
the RW model stops at finite order while that of the CS does not.
The reason is basically due to the need to saturate the zero
modes. This feature did not fail to catch the attention of
Kontsevich, who then pointed out that RW model is an AKSZ model
with parameters and the model gives the characteristic classes of
the holomorphic foliation. We can understand this from the
discussion of section \ref{sec_Flat_Bundle}. In particular, the RW
model is a special case of the model (\ref{act_flat_bunle}). The
'parameters' alluded to in \cite{kontsevich1} are  the harmonic fields in
 (\ref{act_flat_bunle}). At the same time, Kapranov
\cite{kapranov-1997} interpreted the RW model from the point of
view of Atiyah-class. In this section, we will investigate this
model from the field theory perspective and try to endow physical
embodiment to the works \cite{kontsevich1,kapranov-1997}.

RW model is also an AKSZ model. The space of
fields is%
\bea T[1] \Sigma_3 ~\longrightarrow~ T^*[2] T^*[1] M~,\nn\eea%
where $M$ is a HyperK\"ahler manifold\footnote{The construction can be relaxed to the case of holomorphic symplectic manifold.}. The symplectic form for the
target space $T^*[2] T^*[1] M$  is given by%
\bea
 \omega=\delta P_\mu\delta X^\mu+\delta v^\mu\delta q_\mu +\frac{1}{2}\Omega_{ij}\delta X^i\delta X^j~,
 \nn\eea
where $\Omega$ is the holomorphic symplectic 2-form for $M$. We use labels $\mu, \nu, ...$ for real coordinates
 and $i, j, \bar{i}, \bar{j}, ...$ for complex coordinates. The
kinetic term of the AKSZ action is the standard one determined by
the symplectic form above%
\bea S_{kin}=\int\limits_{T[1]\Sigma_3} d^6z\
{\BS{P}}_{\mu}D\BS{X}^{\mu
}+\frac{1}{2}\BS{X}^i\Omega_{ij}D\BS{X}^j+\BS{q}_{\mu
}D\BS{v}^{\mu}~.\label{RW-AKSZ1}\eea%
The interaction term is the one corresponding to the Doubeault
differential%
\bea S_{int}=\int\limits_{T[1]\Sigma_3} d^6z\ \BS{P}_{\bar{i}} \BS{v}^{\bar{i}}~.\label{RW-AKSZ2}\eea%
It is possible to find a LagSubMfld, and restriction of this
action to it gives the RW model \cite{AKSZ_RW}%
\bea S_{RW}&=&\int
g_{i\bar{j}~}dX^{i}_{(0)}\wedge*dX^{\bar{j}}_{(0)}+g_{i\bar{j}}X_{(1)}^{i}\wedge
*d^{\nabla}v^{\bar{j}}-\Omega_{ij}X^{
i}_{(1)}\wedge d^{\nabla}X^{j}_{(1)}\nn\\%
&&-\frac{1}{3} R_{k\bar k\ {j}}^{\ \ \ {
i}}X_{(1)}^{\bar{k}}\wedge\Omega_{{l}{i}} X_{(1)}^{l}\wedge
X^{{j}}_{(1)}v^{\bar{k}}~.\label{RW-BRST}\eea

The first line of the action gives a non-degenerate kinetic term.
This construction of the RW model is of course correct, but when
we are only interested in the computation of invariants of
holomorphic foliation, we can strip down the extraneous parts of
the model and make the geometrical meaning more pronounced.

As far as the RW invariant is concerned, all we need is a cocycle
in the cohomology of Lie algebra of formal Hamiltonian vector
fields  and a Hamiltonian function to plug in. At a given point in
$M$, the holomorphic tangent bundle is identified as
$\mathbb{C}^{2m}$ ($\dim_{\mathbb{R}}M=4m$) and equipped with the
symplectic form $\Omega_{ij}$.

First recall that a tangent vector induces a flow; when we have a
flat connection, we may fix the flow to be the geodesic flow and
unambiguously identify a neighborhood of the origin in the tangent
space of \textsl{x}${}_0$ with a neighborhood of \textsl{x}${}_0$
in $M$. In this way we obtain the so called \emph{normal
coordinates}. Any function in $M$ defined in a neighborhood of
\textsl{x}${}_0$ may be pulled back to the normal coordinate
system. We
denote the geodesic flow induced by the vector $\xi$ as $\exp^*$, so%
\bea
&&X(\textsl{x}_0,\xi)=\textsl{x}_0+\xi^{\mu}-\frac{1}{2}\Gamma^{\mu}_{\nu\rho}\xi^{\nu}\xi^{\rho}
-\frac{1}{6}\xi^{\nu}\xi^{\rho}\xi^{\sigma}\partial_{\nu}\Gamma^{\mu}_{\rho\sigma}
+\frac{1}{3}\xi^{\nu}\xi^{\rho}\xi^{\sigma}\Gamma^{\mu}_{\kappa\nu}\Gamma^{\kappa}_{\rho\sigma}+\cdots\label{Geo_flow}\\%
&&\exp^*\phi(X)=\phi(X(\textsl{x}_0,\xi))=\phi(\textsl{x}_0)+\xi^{\mu}\partial_{\mu}\phi(\textsl{x}_0)
+\frac{1}{2}\xi^{\mu}\xi^{\nu}\nabla_{\mu}\partial_{\nu}\phi(\textsl{x}_0)+\cdots,\nn\eea%
where we assumed that $\Gamma$ is some flat connection.
 Of course, the exponential map requires just a connection, but for the sake of further discussion
  we assume in (\ref{Geo_flow}) that $\Gamma$ is flat.
As a mnemonic, the pull
back of the function $\phi$ by $\exp$ is just the Taylor expansion
of $\phi(\textsl{x}_0+\xi)$ around $\textsl{x}_0$, except that one
uses covariant derivative rather than ordinary derivative. The
same remark applies to the tensors as well, for example%
\bea
\exp^*\phi_{\mu}dX^{\mu}=\phi_{\mu}(\textsl{x}_0)d\xi^{\mu}+\xi^{\nu}\nabla_{\nu}\phi_{\mu}(\textsl{x}_0)d\xi^{\mu}
+\frac{1}{2}\xi^{\nu}\xi^{\rho}\nabla_{\nu}\nabla_{\rho}\phi_{\mu}(\textsl{x}_0)d\xi^{\mu}+\cdots~. \nn\eea

Now we take the base coordinate $\textsl{x}_0$ and also $v$ as
fixed parameters. The target space for the AKSZ model is
\bea {\cal M}=T^{(1,0),\infty}_M(\textsl{x}_0)~,\nn\eea%
where $T^{(1,0),\infty}$ denotes the formal neighborhood of the
zero section of the holomorphic tangent bundle at $\textsl{x}_0$.
Due to the K\"ahler property, the Levi-Civita connection has
either totally holomorphic or totally anti-holomorphic indices,
and the curvature $R_{ij\ \times}^{\ \;\times}=0$, namely so
$\Gamma^i_{jk}$ can be regarded as the flat connection. Of course,
we now only have the holomorphic half of the tangent bundle, so
the geodesic flow is understood as the analytical continuation
away from the real one. For all practical purposes, we just
understand the geodesic flow as given by formal Taylor expansion
as in (\ref{Geo_flow}). Now the target space will be parameterized
by the coordinates $\xi^i$ in the formal neighborhood of
$\textsl{x}_0$. We have originally a holomorphic symplectic form
$dX^i \Omega_{ij} dX^j$,
which we pull back to the $\xi$ coordinate%
\bea
\exp^*dX^i\Omega_{ij}dX^j=d\xi^i\Omega_{ij}d\xi^j+\xi^k(\nabla_k\Omega_{ij})d\xi^id\xi^j+\cdots
=d\xi^i~\Omega_{ij} (\textsl{x}_0)~d\xi^j\label{pull_back_Omega}~.\eea%
We can now set up a free AKSZ theory as in
section \ref{sec_Min_model}. The odd symplectic structure is given by%
\bea\omega=\frac{1}{2}\int\limits_{T[1]\Sigma_3} d^6z\
\delta\BS{\xi}^i~\Omega_{ij}(\textsl{x}_0)~\delta\BS{\xi}^j~.\nn\eea%
We stress that now $\Omega_{ij}$ is \emph{constant} in the $\xi$
space and equal $\Omega_{ij}(\textsl{x}_0)$. The action is%
\bea S=\frac{1}{2}\int\limits_{T[1]\Sigma_3} d^6z\
\BS{\xi}^i~\Omega_{ij}(\textsl{x}_0)~D\BS{\xi}^j~.\label{minimal_model}\eea%
The path integral provides us with the desired cocycle, and we
will evaluate the correlator of the particular function%
\bea
\Theta&=&v^{\bar i}\Theta_{\bar i}=v^{\bar
i}\sum^{\infty}_{n=0}\frac{1}{(n+3)!}R^n_{\bar ii_1\cdots
i_{n+3}}\xi^{i_1}\cdots\xi^{i_{n+3}}~,\nn\\%
\textrm{where}&&R^n_{\bar ii_1\cdots
i_{n+3}}=\nabla_{i_1}\cdots\nabla_{i_n}R_{\bar ii_{n+1}\
i_{n+3}}^{\ \ \ \ \
j}\Omega_{ji_{n+2}}~,\nn\eea%
note that due to the K\"ahler property as well as the covariant
constancy and holomorphy of $\Omega$, all the holomorphic indices
in $R^n$ are symmetric\footnote{Please notice our convention for $R^n_{\bar ii_1\cdots i_{n+3}}$, where
 the superscript $n$ is not a holomorphic index!}. We show in the appendix that%
\bea \bar\partial\Theta+\frac{1}{2}\{\Theta,\Theta\}=0~,\label{Key_id}\eea%
where the Poisson bracket is with respect to  $\Omega$ in $\xi$-direction.

Without specifying what gauge fixing, we still know that the path
integral gives a cocycle with respect to $\Omega$. The correlator is a
function of extra parameters $\textsl{x}_0,v$,
\bea c^q(\BB{X}_\Theta \wedge ... \wedge \BB{X}_\Theta)=f(\textsl{x}_0,v)~,\nn\eea%
which can be regarded as an anti-holomorphic form on $M$. In fact
this is an element in $H^{\bullet}_{\bar\partial}(M)$, for the
Dolbeault differential acts on $c^q$ as the differential in the
graph
cohomology%
\bea \bar{\partial}c^q\big(\BB{X}_\Theta \wedge ... \wedge
\BB{X}_\Theta \big)=-\frac{1}{2}\sum
c^q\big(\BB{X}_\Theta \wedge ... \wedge \BB{X}_{\{\Theta,\Theta\}} \wedge ... \wedge \BB{X}_\Theta \big)\sim
c^q\big(\partial(\BB{X}_\Theta \wedge ... \wedge \BB{X}_\Theta)\big)=0~.\nn\eea%

Furthermore, since the space of $\xi$ now has a flat structure
$\mathbb{C}^{2m}$, we can apply the naive gauge fixing by setting
exact forms to zero as in section \ref{sec_Min_model}. This gauge
fixing automatically keeps only tri-valent graphs. These graphs
are given by contracting the $\xi$'s in $v^{\bar i}R_{\bar ii\
k}^{\ \ l}\Omega_{lj}\xi^i\xi^j\xi^k$ with $\Omega^{-1}$. This
exactly reproduces the Rozansky-Witten invariants. The
$\bar\partial$-closedness is automatic due to $\bar\partial
R\Omega=0$. We pointed out before that the cocycle given by the
path integral does depend on the choice of the LagSubMfld, or the
gauge fixing, it is reasonable to expect that a drastically
different gauge choice shall produce a different class in
$H_{\bar\partial}^{\bullet}$ due to the abundance of cocycles in the
graph cohomology.

The above construction grasps the main feature of Rozansky Witten
model, yet we would like to incorporate also the extra parameters
$\textsl{x}_0, v$ into the theory and furthermore justify the
definition of $\Theta$. In particular we show that RW model fits
the general description of AKSZ model for flat bundles of section
\ref{sec_Flat_Bundle}.

We first complexify $M$ by taking two copies $M\times M$, the
second one is equipped with the opposite complex structure as the
first one. So the diagonal embedding
$M\stackrel{\Delta}{\hookrightarrow}M\times M$ gives the real
slice. The picture here can also be studied in the light of
holomorphic foliation \cite{kontsevich1}. We try to motivate the
analogy between our problem and the holomorphic foliation with a
few words, though this analogy is not strictly necessary for the
rest of the paper, so for the first reading, the reader may jump
over the next five paragraphs.

We label the two copies of $M$ by the holomorphic and
anti-holomorphic coordinate respectively, i.e, a point
$(p,q)\subset M\times M$ is parameterized by%
\bea (X^i(p),X^{\bar i}(q)),\ \ \forall(p,q)\subset M\times M.\nn\eea%
We can take the second factor of the product as the leaf space of
the foliation while the first factor as the transverse direction.
The holomorphic geodesic exponential map amounts to the following
change of coordinates%
\bea(X^i(p),X^{\bar i}(q))\rightarrow(\exp_{\xi}X^i(q),X^{\bar i}(q)).\label{x_0_and_X_bar}\eea%

For a foliation with constant co-dimension, we have principle
bundle structure. The fiber is (rather abstractly) all possible
ways of identifying the transverse space at the neighborhood of a
point with the Euclidean space $\mathbb{C}^{2m}$. The structure
group is by definition isomorphic to the fiber and can be taken as
the group of formal diffeomorphism of the transverse direction.
Now that there is a symplectic structure $\Omega_{ij}$ in the
transverse direction, the relevant structure group should become
the group of formal symplectomorphism. The previously defined
exponential map offers one way of such identification, namely at
point $q$, the point $p$ in the transverse space is identified
with $\xi\in\BB{C}^{2n}$ through $X^i(p)=\exp_{\xi}X^i(q)$. So the
exponential map is a section of the principle bundle.

This principle has a flat connection. Here we follow the work of
Fuks \cite{FuksI}. Let the foliation be determined by a system of
2n 1-forms $\theta^i$, which means the leaf is the null-space of
these forms. By the Frobenius theorem for an integrable system of
1-forms, the differential $d\theta^i$ is%
\bea d\theta^i=\gamma^i_{\ j}\wedge\theta^j~,\nn\eea%
where $\gamma$ is again some 1-forms, and can be thought of as the
connection of the principle bundle. This connection is flat only
along the leaf: $(d\gamma^i_{\ j}-\gamma^i_{\ k}\gamma^k_{\
j})\theta^j=0$. But this in turn implies $d\gamma^i_{\
j}-\gamma^i_{\ k}\gamma^k_{\ j}=\gamma^i_{\ jk}\theta^k$ for some
1-form $\gamma^i_{\ jk}$. One can carry on this procedure and
obtain a collection of such $\gamma$'s, and out of these one can
construct a connection%
\bea
\Gamma=\Gamma^i\frac{\partial}{\partial{\eta^i}}=\big(\theta^i+\gamma^i_j\eta^j+\gamma^i_{jk}\eta^j\eta^k+\cdots\big)\frac{\partial}{\partial{\eta^i}},\nn\eea
where $\eta$ is some formal variable. This connection now takes
value in the Lie algebra of formal vector fields in the
$\eta$-space. This connection is flat in all directions.

Next we apply the above machinery to bear upon our problem. Due to
the mapping Eq.\ref{x_0_and_X_bar}, the previous $\textsl{x}_0(q)$
is identified as $X^i(q)=\big(X^{\bar i}(q)\big)^*$. The
holomorphic foliation is determined by $2n$ 1-forms $dX^i$,
because the leaf is clearly the null space of these 1-forms. It
will be shown in the appendix that the pull back of these 1-forms
$\exp_{\xi}^*dX^i$ is the linear combination of the system
$\theta^i=d\xi^i-dX^{\bar i}\{\Theta_{\bar i},\xi^i\}$. According
to the above recipe, we
differentiate the 1-forms and it turns out that%
\bea \gamma^i_{\ j}\sim dX^{\bar i}\partial_{\xi^j}\{\Theta_{\bar
i},\xi^i\};\ \gamma^i_{\ jk}\sim dX^{\bar
i}\partial_{\xi^j}\partial_{\xi^k}\{\Theta_{\bar i},\xi^i\};\ \cdots~.\nn\eea%
And the flat connection for the foliation is given by%
\bea &&\Gamma=\Big(d\xi^i-dX^{\bar
i}\sum_{n=0}^{n=\infty}\frac{1}{n!}(\eta\partial_{\xi})^n\{\Theta_{\bar
i},\xi^i\}\Big)\partial_{\eta^i}\nn\\%
&&d\Gamma-\Gamma\Gamma=0\nn\eea%
This connection is flat in all (including the $\xi$) directions
after applying Eq.\ref{Key_id}.

Now that we have a flat principle bundle defined over $M\times M$,
but we can restrict it to the diagonal (given by $\xi=0$). The
pull back of the connection, for which we use the same symbol, is%
\bea \Gamma=-dX^{\bar
i}\sum_{n=0}^{n=\infty}\frac{1}{n!}(\eta\partial_{\xi})^n\{\Theta_{\bar
i},\xi^i\}\partial_{\eta^i}\Big|_{\xi=0}\nn\eea%
This is obviously is just $dX^{\bar i}\{\Theta_{\bar i},\xi^i\}$
with all the $\xi$'s replaced with $\eta$. Finally, we see that it
is in this way Eq.\ref{Key_id} is interpreted as the flat
connection of holomorphic foliation and the correlator of the RW
model gives rise to the characteristic class of the holomorphic
foliation.

Back from our digression, the key observation by Kapranov
\cite{kapranov-1997} is the following, under the mapping
$T^{(1,0),\infty}\stackrel{\exp}{\rightarrow}M\times M$,
$(\xi,X^{\bar i})\rightarrow(\exp_{\xi}(X^{\bar i})^*,X^{\bar
i})$, the differential $\bar\partial$ is pulled by as%
\bea&&\big(\{\Theta_{\bar i},\cdot\},\bar\partial_{\bar
i}\big)\stackrel{\exp}{\rightarrow} \big(0,\bar\partial_{\bar
i}\big)~.\label{Kapranov}\eea%
This motivates (\ref{Key_id}) because the Dolbeault differential
on the rhs is nilpotent. It is also instructive to prove this
relation explicitly which we do in the appendix.

On the space $M\times M$, we can construct the GrMfld%
\bea {\cal M}=M\times T^{(0,1)}[2]T^{(0,1)}[1]M\nn\eea
parameterized by $(X^i,P_{\bar i},q_{\bar i},v^{\bar i},X^{\bar
i})$. It has the even symplectic form%
\bea \omega=\delta P_{\bar i}\wedge\delta X^{\bar i}+\delta
q_{\bar i}\wedge\delta v^{\bar i}+\frac{1}{2}\Omega_{ij}\delta
X^i\wedge\delta X^j~.\label{symp111}\eea%
With this data one can set up the standard AKSZ model, whose
homological vector field is $\bar\partial$--the rhs of
Eq.\ref{Kapranov},%
\bea &&S=\int d^6z\ \BS{P}_{\bar i}D\BS{X}^{\bar i}+\BS{q}_{\bar
i}D\BS{v}^{\bar
i}+\frac{1}{2}\BS{X}^i\Omega_{ij}D\BS{X}^{j}+\BS{P}_{\bar
i}\BS{v}^{\bar i}~.\label{previous}\eea
 This action is a truncation of the AKSZ action (\ref{RW-AKSZ1})+(\ref{RW-AKSZ2}) and it still
  reduces to the gauge fixed RW action (\ref{RW-BRST}) along the lines presented in \cite{AKSZ_RW}.

While for the manifold $T^{(1,0)\infty}_M$ one has the GrMfld%
\bea {\cal M}=T^{(1,0)\infty}M\oplus
T^{(0,1)}[2]T^{(0,1)}[1]M\nn\eea%
parameterized by $(\xi^i,P_{\bar i},q_{\bar i},v^{\bar i},X^{\bar
i})$. We want to pull the model
 (\ref{previous}) on $M\times T^{(0,1)}[2]T^{(0,1)}[1]M$ back to
this target space. The required change of variable is
($\textsl{x}_0=(X^{\bar i})^*$) \bea
\xi^i&\stackrel{\exp}{\rightarrow}&\textsl{x}^i_0+\xi^i-\frac{1}{2}\Gamma^i_{jk}\xi^j\xi^k+\cdots~
=e^{\xi^i\partial_{\textsl{x}^i_0}-\xi^i\xi^j\Gamma^k_{ij}\partial_{\xi^k}}\textsl{x}_0^i,\nn\\
P_{\bar i},q_{\bar i},v^{\bar i},X^{\bar i}&\rightarrow&P_{\bar
i},q_{\bar i},v^{\bar i},X^{\bar i}~.\nn\eea
The same calculation that led to (\ref{Kapranov}) shows%
\bea \delta X^i=\big(\delta \xi^j-\delta X^{\bar i}\{\Theta_{\bar
i},\xi^j\})\frac{\partial X^i(\xi)}{\partial\xi^j}~.\nn\eea
The holomorphic symplectic form is pulled back according to (using
 the equation (\ref{pull_back_Omega}))%
\bea \exp^*\delta X^i\Omega_{ij}\delta X^j&=&(\delta\xi^i-\delta
X^{\bar i}\{\Theta_{\bar
i},\xi^i\})(\exp_{\xi}^*\Omega_{ij})(\delta\xi^j-\delta
X^{\bar j}\{\Theta_{\bar j},\xi^j\})\nn\\%
&=&\delta\xi^i\Omega_{ij}\delta\xi^j-2\delta X^{\bar
i}\{\Theta_{\bar i},\xi^i\}\Omega_{ij}\delta \xi^j+\delta X^{\bar
i}\{\Theta_{\bar i},\xi^i\}\Omega_{ij}\delta X^{\bar
j}\{\Theta_{\bar j},\xi^j\}\nn\\%
&=&\delta\xi^i\Omega_{ij}\delta\xi^j-2\delta X^{\bar
i}\frac{\partial\Theta_{\bar i}}{\partial\xi^j}\delta \xi^j-\delta
X^{\bar i}\delta X^{\bar j}\{\Theta_{\bar i},\Theta_{\bar j}\}\nn\\%
&=&\delta\xi^i\Omega_{ij}\delta\xi^j-2\delta X^{\bar i}\delta
\Theta_{\bar i}~.\nn\eea
So the symplectic form Eq.\ref{symp111} is pulled back to%
\bea\exp^*\omega=\int d^6z\ \delta\BS{P}_{\bar
i}\wedge\delta\BS{X}^{\bar i}+\delta\BS{q}_{\bar
i}\delta\BS{v}^{\bar
i}+\frac{1}{2}\delta\BS{\xi}^i\Omega_{ij}\delta\BS{\xi}^j-\delta\BS{X}^{\bar
i}\delta\BS{\Theta}_{\bar i}~.\nn\eea%
The action Eq.\ref{previous} is pulled back as%
\bea &&\exp^*S=\int d^6z\ (\BS{P}_{\bar i}+\BS{\Theta}_{\bar
i})D\BS{X}^{\bar
i}+\frac{1}{2}\BS{\xi}^i\Omega_{ij}D\BS{\xi}^j+\BS{q}_{\bar
i}D\BS{v}^{\bar i}+\BS{P}_{\bar
i}\BS{v}^{\bar i}~.\nn\eea%
Since now the momentum dual to $X^{\bar i}$ is $P_{\bar
i}+\Theta_{\bar i}$, it is proper that we changed variable
$\tilde{P}_{\bar i}:=P_{\bar i}+\Theta_{\bar i}$%
\bea &&\exp^*S=\int d^6z\ \tilde{\BS{P}}_{\bar i}D\BS{X}^{\bar
i}+\frac{1}{2}\BS{\xi}^i\Omega_{ij}D\BS{\xi}^j+\BS{q}_{\bar
i}D\BS{v}^{\bar i}+\tilde{\BS{P}}_{\bar i}\BS{v}^{\bar
i}-\BS{\Theta_{\bar i}}\BS{v}^{\bar i}\nn\eea%
Note that $\tilde P_{\bar i}v^{\bar i}-\Theta_{\bar i}v^{\bar i}$
generates the vector field $-(v^{\bar i}\bar\partial_{\bar
i}+v^{\bar i}\{\Theta_{\bar i},\cdot\})$ on functions of $X^{\bar
i}$ and $\xi^i$, which is the lhs of Eq.\ref{Kapranov}. This model
is totally in line with the general picture for the model given by
 (\ref{act_flat_bunle}).

We can perform the partial gauge fixing in the $\tilde P_{\bar
i},X^{\bar i},q_{\bar i},v^{\bar i}$ sector by setting $\tilde
P_{\bar i}=q_{\bar i}=0$. Then we are left with the action with
only $\xi$ as dynamical variables, relegating $X_{\bar i}$ and
$v^{\bar i}$ as extra parameters. This gives back the odd Chern-Simons
model (\ref{minimal_model}) with an interaction term $\Theta$.

\section{Summary}
\label{summary}

In this paper, we have explained the idea of using AKSZ-BV path
integral as a construction of cocycles and that its relation to
the graph cohomology is nothing but the Feynman integrals and the
standard Wick's theorem. We took the construction of
\cite{Hamilton:2007tg} and put it into a concrete physical system.
In particular, we discussed how to deal with zero modes which must
exist in any realistic field theory. This leads to the embodiment
of Kontsevich's idea of applying homomorphism to a cocycle of Lie
algebra cohomology to obtain secondary Chern-Simons type
invariants (characteristic classes of flat bundles).
  Thus we conclude that the AKSZ construction of TFT is powerful not
  only at the classical level, it also offers very unified perturbative treatment of the corresponding
   TFTs.

We constructed
the odd Chern-Simons theory over the target $\mathbb{R}^{2n}$ and showed
that its perturbation expansion is identical to that of the
Chern-Simons theory, in particular, we obtained identical weight
function for each given graph.
We did this for the
Rozansky-Witten model painstakingly, and showed from the field
theory perspective that this model fits the picture painted by
Kontsevich, namely, it is a model associated with a flat bundle
related to the holomorphic foliation.

 The further issues include of course applying the presented ideas for the
general AKSZ model for different algebroids and foliations and
construct explicitly characteristic classes and invariants.
 One can naturally associate  3D AKSZ models to Courant algebroids and Lie algebroids.
   Thus applying the ideas presented in this work one may hope to
    obtain interesting characteristic classes for these algebroids.  The main complication in
     the treatment of these models would be the application of the exponential map carefully, or in
      other words, performing  the covariant Taylor expansions.

      Another interesting issue would to apply the formal BV arguments from section \ref{sec_BV}
       to a wide class of quantum observables. In this paper we concentrate our attention on observables
        which are written as full integral over source manifold. The BV algebra of these observales can be
         mapped to the algebra of functions on the target space. However we may look at the quantum observables
          which are integrals over cycles on the source (or even full Wilson loops).  One has to embed these wide
           class of observables into the BV framework and calculate the corresponding BV algebra generated by
            them. The path integral evaluations of those observables should still give rise to a cocycle for
             some Lie algebra. We hope to return to this idea in the future.

\bigskip\bigskip

\noindent{\bf\Large Acknowledgement}:
\bigskip

\noindent
 The research of M.Z. was
supported  by VR-grant 621-2008-4273. The authors would like to
thank Francesco Bonechi for helpful discussions.

\bigskip\bigskip

\appendix
\section{Brackets of Even and Odd Type}

In this section we fix the sign conventions of the symplectic
form, Poisson bracket and odd Laplacian etc. These signs are important
for the perturbation theory, it is worth the effort.

The degree $n$ symplectic form%
\bea \Omega=\sum_{A<B}\Omega_{AB}~dX^A\wedge dX^B~,~~~~|A|+ |B|=n~,\nn\eea%
 where $|A|= \deg X^A$ and $|B|=\deg X^B$.
We always assume that $\Omega_{AB}$ is a constant for simplicity,
and it satisfies $\Omega_{AB}=(-1)^{(|A|+1)(|B|+1)}\Omega_{BA}$,
matching the graded commutativity
$dX^AdX^B=(-1)^{(|A|+1)(|B|+1)}dX^BdX^A$.

We take the odd symplectic form as the starting point. This gives
rise to an odd Poisson bracket, which can be induced from the odd
Laplacian according to the formula (\ref{LAP_BV_GR_BR}). We fix the
convention for this odd Laplacian first, and from there we
\emph{derive} the conventions of other brackets. The reason is
that whether or not a Laplacian annihilates some function is
crucial for the discussion of section \ref{sec_A_3D_Model}. We assume
that $\Omega_{AB}$ is a constant for simplicity, and fix the
following
\bea \sum_{A<B}\Omega_{AB}~dX^A\wedge dX^B\Rightarrow
\Delta=\sum_{A<B}(\Omega^{-1})^{AB}~\frac{\partial}{\partial
X^A}\frac{\partial}{\partial X^B}~.\label{rule_Lap}\eea
One may check that $\Delta$ satisfies (\ref{LAP_BV_GR_BR}), with
the Poisson bracket given by%
\bea\{f,g\}=\sum_{A,B}(\Omega^{-1})^{AB}(f\overleftarrow{\partial_A}
)\partial_Bg ~.\label{Poisson}\eea%
Note that our convention for the right derivative is
$X^B\overleftarrow{\partial_A}=(-1)^{|A|} \delta^B_A$.

 The bracket (\ref{Poisson}) is derived for the odd case, but we take this as
the definition of the Poisson bracket, both for even and odd. This
bracket satisfies $\{f,g\}=-(-1)^{(f+n)(g+n)}\{g,f\}$.

Suppose we have now a deg $n$ symplectic GrMfld ${\cal M}$, with
symplectic form $\Omega$. Because we give degree 1 to $\delta$, as
a result, $\delta\theta=-\theta\delta$ and we dispense with the
$\wedge$. In the AKSZ construction, we build a TFT on a dimension
$n+1$ source manifold $\Sigma_{n+1}$ with ${\cal M}$ as the
target. For such degree $n$ GrMfld, we can form the degree $-1$
symplectic
form over Maps$(T[1]\Sigma_{n+1},{\cal M})$ by%
\bea \omega=\int\limits_{T[1]\Sigma_{n+1}} d^{2(n+1)}z~\big
(\sum_{A<B}\Omega_{AB}\delta{\BS X^A(z)}\delta{\BS X^B(z)}\big)~,\label{sym33838}\eea%
where $d^{2(n+1)}z=d^{n+1}\xi~d^{n+1}\theta$ and $\BS X$ stands for a
map from $T[1]\Sigma_{n+1}$ to ${\cal M}$.
 This form has the desired
degree $-1$ because the measure carries degree $-(n+1)$.

We may obtain the odd symplectic form written in component fields
by integrating out $d^{n+1}\theta$. Assume $\Omega_{AB}$ is
a constant the symplectic form (\ref{sym33838}) can be rewritten as
\bea
 \omega=\sum_p(-1)^{A(n+1-p)+p}\int d^{n+1}\xi~
\Omega_{AB}\delta{X}^A_{(p)} \delta{X}_{(n+1-p)}^B~.\nn\eea
The Laplacian according to the rule (\ref{rule_Lap}) is
\bea \Delta=\sum_{p,A<B}(-1)^{A(n+1-p)+p+(A+1)(B+1)}\int d^{n+1}\xi~
(\Omega^{-1})^{AB}\frac{\delta}{\delta
X^A_{(p)}(\xi)}\frac{\delta}{\delta X_{(n+1-p)}^B(\xi)}~.\nn\eea
This naive form of the Laplacian must be improved, otherwise, when it hits a local functional, it will produce $\delta(0)$. Nevertheless, if we proceed and investigate $\Delta\int d^{2(n+1)}z~ f(\BS{X}(x))=0$, we find
 the following formal expression
\bea
&& \Delta \int d^{2(n+1)}z~ f(\BS X(z)) \nn \\
&&= 4\sum_p \binom{n+1}{p}
(-1)^p\sum_{A<B}(-1)^{|A||B|+n+1}(\Omega^{-1})^{AB}\int d^{n+1}\xi~
\partial_{B}\partial_Af(X(\xi)) \delta(0)~.\nn\eea%
Note that this sum formally vanishes for any $n+1$,
$$\sum\limits_{p=0}^{n+1} \binom{n+1}{p} (-1)^{p}=(1-1)^{n+1}=0~.$$
A better definition of $\Delta$ with regularization\footnote{We use the regularization of odd Laplacian which is
 similar to the one discussed in \cite{Costello:2007ei}.}, which is
suited both for separating the subtlety from zero modes and for
renormalization is the following. We expand any $p$-form on the
source
manifold into eigen-modes of the self-adjoint operator $\square=\{d^{\dagger}, d\}$,
 where $d$ is de Rham differential and $d^{\dagger}$ is its adjoint,
\bea &&X_{(p)}=X_{I_p}\psi^{I_p}~;\nn\\%
&&\square\psi^{I_p}=\lambda_{I_p}^2\psi^{I_p}~;\nn\\%
 &&\sum_{I_p}\psi^{I_p}(x)_{i_1\cdots i_p}(*\psi^{I_p}(y))_{i_{p+1}\cdots i_{n+1}}=\epsilon_{i_1\cdots i_{n+1}}\delta(x-y)~.\nn\eea%
After changing variables from $X^A_{(p)}$ to $X^A_{I_p}$, the original Laplacian becomes%
\bea
\Delta=\sum_{A<B,p,I_p}(-1)^{Ap+p(n+1)+AB}(\Omega^{-1})^{AB}~\frac{\partial}{\partial
X^A_{I_{n+1-p}}}\frac{\partial}{\partial
X^B_{I_p}}~,\label{Laplacian_mode}\nn\eea To regularize this
expression, one inserts the factor
$\exp{(-\epsilon^2\lambda^2_{I_p})}$ in the summation. This
regularization is commonly known as the heat kernel
regularization, we denote%
\bea
\Delta_{\epsilon}=\sum_{A<B,p,I_p}(-1)^{Ap+p(n+1)+AB}e^{(-\epsilon^2\lambda^2_{I_p})}(\Omega^{-1})^{AB}~\frac{\partial}{\partial
X^A_{I_{n+1-p}}}\frac{\partial}{\partial X^B_{I_{p}}}~,\nn\eea%
What happens here is that we are effectively replacing the original
Laplacian with%
\bea &&\Delta_{\epsilon}=\sum_{p,A<B}(-1)^{A(n+1-p)+p+AB+n+1}\int
d^{n+1}\xi_1d^{n+1}\xi_2 ~K_p(\xi_1,\xi_2,\epsilon)~\frac{\delta}{\delta
X^A_{(p)}(\xi_1)}\frac{\delta}{\delta X_{(n+1-p)}^B(\xi_2)}~,\nn\\%
&&K_p(\xi_1,\xi_2,\epsilon)=\sum_{I_p}e^{-\epsilon^2\lambda_{I_p}^2}\psi^{I_p}(\xi_1)\wedge\psi^{I_{n+1-p}}(\xi_2)~.\nn\eea
For small $\epsilon$, the heat kernel $K$ asymptotes to%
\bea K_p(\xi_1,\xi_2,\epsilon)\sim
\epsilon^{-(n+1)}e^{\frac{|\xi_1-\xi_2|^2}{4\epsilon^2}}~,\nn\eea%
which is just the smeared delta function, and we recover the
original definition of $\Delta$ in the limit $\epsilon \rightarrow 0$.

It turns out that $\Delta_{\epsilon}$ acts on a full integral as%
\bea \Delta_{\epsilon}\int d^{2(n+1)}z~f(\BS
X(z))=\sum_{p,A<B}(-1)^p(\Omega^{-1})^{AB}\int d^{n+1}\xi~
K_p(\xi,\xi,\epsilon)\partial_A\partial_Bf(X(\xi))~.\nn\eea%
Suppose that $\partial_A\partial_B f(X(\xi))$ is a constant, then
the sum over $p$ gives nothing but the index of the de Rham
operator on $\Sigma_{n+1}$
\bea \sum_p(-1)^p\int d^{n+1}\xi~
K_p(\xi,\xi,\epsilon)=\Tr[e^{-\epsilon^2\square}(-1)^p]=\chi(\Sigma_{n+1})~.\nn\eea
While for odd $(n+1)$ which is the main interest of this paper, the
sum can be reshuffled into%
\bea
\Delta_{\epsilon}\int d^{2(n+1)}z~f(\BS
X(z))=\sum_{p\leq(n/2),A,B}(-1)^p(\Omega^{-1})^{AB}\int d^{n+1}\xi~
K_p(\xi,\xi,\epsilon)\partial_A\partial_Bf(X(\xi))~,\label{ddkdk3303}\eea
  where now sum is taken over all $A$ and $B$. We used the fact that
   if $\psi^{I_p}$ is eigenfunction of $\square$ with the eigenvalue $\lambda_{I_p}^2$,
    then $*\psi^{I_p}$ is eigenfunction of $\square$ with the same eigenvalue.
 The expression (\ref{ddkdk3303})   vanishes for nonzero $\epsilon$ since
 $(\Omega^{-1})^{AB} \partial_A \partial_B f=0$ due to contraction of (graded)symmetric
  with (graded)anti-symmetric and this happens only when $n+1$ is even.
 Here we can see the crucial difference between
even and odd dimension theory.

We would now like to investigate the relation between the Poisson
bracket on the target space ${\cal M}$ and the induced odd bracket
in the mapping space, in particular whether $\Delta(\int f\int
g)=\int \{f,g\}$ is true. After a lengthy but straightforward
calculation, we obtain%
\bea \Delta\big({\Large\textrm{$\smallint$}}d^{2(n+1)}z_1\ f(\BS
X(z_1)){\Large\textrm{$\smallint$}}d^{2(n+1)}z_2\ g(\BS
X(z_2))\big)=(-1)^{f}{\Large\textrm{$\smallint$}}d^{2(n+1)}z~\{f(\BS
X(z)),g(\BS X(z))\}~.\nn\eea%
From this we can get%
\bea \big\{{\Large\textrm{$\smallint$}}d^{2(n+1)}z\ f(\BS
X(z)),{\Large\textrm{$\smallint$}} d^{2(n+1)}z\ g(\BS
X(z))\big\}=(-1)^{n+1}{\Large\textrm{$\smallint$}} d^{2(n+1)}z\
\{f(\BS X(z)),g(\BS X(z))\}~.\nn\eea

Furthermore, in the text we quite often treat the harmonic modes (i.e., the eigenfunctions
 with $\lambda_{I_p}^2=0$)
as parameters of the theory and the path integral is taken only
for the non-harmonic fields. If we denote the Laplacian of the
non-harmonic fields by $\Delta'$, the Ward identity is given by
\bea \int\limits_{{\cal L}}\Delta'(\cdots)=0~,\nn\eea%
 where ${\cal L}$ is Lagrangian only in non-harmonic sector.
Thus we should investigate what is the bracket induced by the
$\Delta'$.

To do this, we do not need the regularization above. We restrict
ourselves to the case of rational homology sphere for simplicity.
Then in the mode sum of (\ref{Laplacian_mode}), we need to
exclude two modes%
\bea \psi^h_0=\frac{1}{\sqrt{vol}};\ \psi^h_{n+1}=\frac{\sqrt
g}{\sqrt{vol}}\epsilon_{i_1\cdots i_{n+1}}~.\nn\eea
As a result we obtain%
\bea &&\Delta'\big({\Large\textrm{$\smallint$}}d^{2(n+1)}z_1\
f(\BS X(z_1)){\Large\textrm{$\smallint$}}d^{2(n+1)}z_2\ g(\BS
X(z_2))\big)=(-1)^{f}{\Large\textrm{$\smallint$}}
d^{2(n+1)}z~\{f(\BS X(z)),g(\BS X(z))\}\nn\\%
&&-\sum_{A,B}(\Omega^{-1})^{AB}(-1)^{nf+A(n+1)}(\frac{1}{\sqrt{vol}}{\Large\textrm{$\smallint$}}\psi^h_{n+1}(\xi)f(X(\xi))\overleftarrow{\partial}_A)
{\Large\textrm{$\smallint$}}
d^{2(n+1)}z\partial_Bg(\BS X(z))\nn\\%
&&-\sum_{A,B}(\Omega^{-1})^{AB}(-1)^{f(n+1)+g+gf+AB+n+1}(\frac{1}{\sqrt{vol}}{\Large\textrm{$\smallint$}}
\psi^h_{n+1}(\xi)g(X(\xi))\overleftarrow{\partial}_A){\Large\textrm{$\smallint$}}
d^{2(n+1)}z\partial_Bf(\BS X(z))~.\nn\eea%
The first term is the usual one, while the last two are due to the
exclusion of the zero modes.

To make sense of the formula, we have to resort to the explicit
gauge fixing and the Feynmann rules
we introduced in section \ref{sec_gauge_fix}. Thus from now on
$n+1=3$. Our claim
is that we
can do the following replacement in the path integral%
\bea\int\limits_{{\cal L}}
(\frac{1}{\sqrt{vol}}{\Large\textrm{$\smallint$}}\psi^h_3(\xi)f(X(\xi))\cdots=f(\textsl{x}_0)\int\limits_{{\cal
L}}\cdots,\nn\eea%
\emph{as long as} all the other insertions are of the form $\int
d^6z~g(\BS{X})$. The reasoning is, suppose that $\cdots$ consists
of $q$ insertions, then the number of propagators routed amongst
themselves is $\#=3q/2$. If the insertion $\int
d^3\xi~\psi^h_3(\xi)f(X(\xi))$ was connected to the rest of the
diagram with $p\geq 2$ propagators, then $\#$ will be reduced to
$\#\leq(q-2p)/3$, forcing $2$-valent vertex to appear somewhere and
 it vanishes within our rules.

With this consideration, the second term of the previous formula
under the path integral becomes
\bea
-(-1)^{nf}\big\{f(\textsl{x}_0),\int\limits_{{\cal
L}}\cdots\big\}~,\nn
\eea
 where the Poisson bracket is now taken over
$\textsl{x}_0$. The path integral over the non-harmonic fields
produces a function of $\textsl{x}_0$, 
accordingly the path integral should be interpreted now as a
cochain of the CE complex of formal Hamiltonian vector fields of
${\cal M}$, \emph{taking values in $C^{\infty}({\cal M})$}. So the
differential of such cochains must be modified correspondingly,
and this new differential is
induced by $\Delta'$.%
\bea&&(\delta
c^q)(\BB{X}_{f_0}\wedge\cdots\BB{X}_{f_{q+1}})\nn\\%
&=&\sum_{i<j}(-1)^{f_i+(f_i+n+1)(f_0+\cdots
f_{i-1}+i(n+1))+(f_j+n+1)(f_0+\cdots
f_{j-1}+j(n+1))+(f_i+n+1)(f_j+n+1)}\times\nn\\%
&&c^q(\BB{X}_{\{f_i,f_j\}}\wedge\cdots\widehat{\BB{X}}_{f_i}
\wedge\cdots\widehat{\BB{X}}_{f_j}\wedge\cdots\BB{X}_{f_{q+1}})\label{general_diff}\\%
&&-\sum_{i}(-1)^{f_in+(f_i+n+1)(f_0+\cdots f_{i-1}+i(n+1))+(f_i+n)\deg
c^q}\big\{f_i(\textsl{x}_0),
c^q(\BB{X}_{f_0}\wedge\cdots\widehat{\BB{X}}_{f_i}
\wedge\cdots\BB{X}_{f_{q+1}})\big\}~.\nn\eea%
In fact, this formula is completely in accordance with the de Rham
differential%
\bea
d\omega(X_0,\cdots,X_q)&=&\sum_{i<j}(-1)^{i+j+1}\omega([X_i,X_j],X_0,\cdots,\hat{X}_i,\cdots,\hat{X}_j,\cdots,X_q)\nn\\
&&-\sum_i(-1)^iX_i\omega(X_0,\cdots\hat{X}_i,\cdots X_q).\nn\eea
Because of the formal relation $\int\limits_{{\cal
L}}\Delta'\cdots=0$, the path integral is a cocycle for the
modified differential. What is not expected in
(\ref{general_diff}) is perhaps the sudden appearance of
$(f_i+n)\deg c^q$. When one derives the formula of the
differential using $\Delta'$, one is restricted to $n+1=3$ and
$q=2k-1$, so the factor $(f_i+n)\deg c^q=(f_i+2)6k$ is invisible.
But for a general CE differential for general degree $n$ bracket,
this factor is needed in order $\delta^2=0$. If one wishes to
check this point for himself, he will find the following useful
\bea &&\{f,g\}=-(-1)^{(f+n)(g+n)}\{g,f\}~,\nn\\%
&&\{\{f,g\},h\}+\{\{g,h\},f\}(-1)^{(g+h)(f+n)}+\{\{h,f\},g\}(-1)^{(g+f)(h+n)}=0~.\nn\eea

Lastly, we also check that for an odd symplectic form, the
Hamiltonian vector fields satisfy%
\bea
[\BB{X}_f,\BB{X}_g] \equiv \BB{X}_f\BB{X}_g-(-1)^{(|f|+1)(|g|+1)}\BB{X}_g\BB{X}_f=\BB{X}_{\{f,g\}}~.\nn\eea%
We do this in Darboux coordinates for simplicity. The symplectic
form is $\omega=\delta x^+_a\delta x^a$ where $x^+$ is odd $x$ is
even\bea
[\BB{X}_f,\BB{X}_g]&=&\{f,g\frac{\overleftarrow{\partial}}{\partial
x_a^+}\}\frac{\partial}{\partial
x^a}+\{f,g\frac{\overleftarrow{\partial}}{\partial
x^a}\}\frac{\partial}{\partial
x^+_a}-(-1)^{(|f|+1)(|g|+1)}(f\leftrightarrow g)\nn\\%
&=&\BB{X}_{\{f,g\}}+(-1)^{|g|}\{f\frac{\overleftarrow{\partial}}{\partial
x_a^+},g\}\frac{\partial}{\partial
x^a}-\{f\frac{\overleftarrow{\partial}}{\partial
x^a},g\}\frac{\partial}{\partial x^+_a}\nn\\%
&&-(-1)^{(|f|+1)(|g|+1)}
\{g,f\frac{\overleftarrow{\partial}}{\partial
x_a^+}\}\frac{\partial}{\partial x^a}
-(-1)^{(|f|+1)(|g|+1)}\{g,f\frac{\overleftarrow{\partial}}{\partial
x^a}\}\frac{\partial}{\partial x^+_a}=\BB{X}_{\{f,g\}}~,\nn\eea%
where we have used $\{f,g\}=-(-1)^{(|f|+1)(|g|+1)}\{g,f\}$.

\section{$L_{\infty}$ Structure from HyperK\"ahler Manifold}

In this Appendix we present some explicit formulas about $L_\infty$-structure  for the hyperK\"ahler manifold.
 The idea was presented in \cite{kapranov-1997}, but we could not read off the explicit numerical factors from this work. Therefore we present
  our own derivation of these relations.
 All expressions are written in complex coordinates.

For a hyperK\"ahler manifold, the three indices $i,j,k$ are
totally symmetric in $(R\Omega)_{\bar iijk}=R_{\bar ii\ k}^{\ \
l}\Omega_{lj}$ and $\partial_{[\bar j}(R\Omega)_{\bar i]ijk}=0$.
If we define%
\bea R^n_{\bar il_1\cdots
l_{n+3}} \equiv \nabla_{l_1}\cdots\nabla_{l_n}(R\Omega)_{\bar
il_{n+1}l_{n+2}l_{n+3}}~,\nn\eea%
then apply the covariant derivatives to $0=\partial_{[\bar
j}(R\Omega)_{\bar i]ijk}$%
\bea 0&=&\frac{1}{(n+3)!}\nabla_{l_1\cdots l_n}\bar\partial_{[\bar
j}(R\Omega)_{\bar
i]l_{n+1}l_{n+2}l_{n+3}}+\textrm{perm in }l_i\nn\\%
&=&\sum_{k=1}^{n-1}\frac{(n-k+2)}{(n+3)!}\nabla_{l_1\cdots
l_k}\big[(R_{[\bar j|l_{k+1}\
l_{k+2}}^{\hspace{.75cm}m})R^{n-k-1}_{\bar i]m\cdots
l_{n+3}}\big]+\frac{1}{(n+3)!}\bar\partial_{[\bar j}R^n_{\bar
i]l_1\cdots l_{n+3}}+\textrm{perm in }l_i~.\nn\eea%
So we get%
\bea \bar\partial_{[\bar j}R^n_{\bar i]l_1\cdots
l_{n+3}}=-\sum_{k=0}^{n-1}\frac{(n-k+2)}{(n+3)!}\nabla_{l_1\cdots
l_k}\big[(R_{[\bar j|l_{k+1}\
l_{k+2}}^{\hspace{.75cm}m})R^{n-k-1}_{\bar i]m\cdots
l_{n+3}}\big]+\textrm{perm in }l_i~.\nn\eea%
The rhs can be worked out explicitly%
\bea\textrm{rhs}&=&\sum_{k=0}^{n-1}\sum_{p=0}^{k}\binom{k}{p}\frac{(n-k+2)}{(n+3)!}\nabla_{l_1\cdots
l_p}(R_{[\bar j|l_{k+1}\
l_{k+2}}^{\hspace{.75cm}m})\nabla_{l_{p+1}\cdots
l_{k}}R^{n-k-1}_{\bar i]m\cdots l_{n+3}}+\textrm{perm in }l_i\nn\\%
&=&\sum_{k=0}^{n-1}\sum_{p=0}^{k} \binom{k}{p}\frac{(n-k+2)}{(n+3)!}R^p_{[\bar
j|l_1\cdots l_pl_{k+1}l_{k+2}m}\Omega^{mn}R^{n-p-1}_{\bar
i]l_{p+1}\cdots l_{k}n\cdots l_{n+3}}+\textrm{perm in }l_i\nn\\%
&=&\sum_{p=0}^{n-1}\sum_{k=p}^{n-1}\frac{k!(n-k+2)}{(n+3)!p!(k-p)!}R^p_{[\bar
j|l_1\cdots l_{p+2}m}\Omega^{mn}R^{n-p-1}_{\bar i]l_{p+3}\cdots
l_{n+3}m}+\textrm{perm in }l_i~.\nn\eea
The summation of factorials can be worked out as follows%
\bea
\sum_{k=p}^n\frac{k!}{(k-p)!}&=&\lim_{\epsilon\rightarrow0}\sum_{k=p}^n\frac{k!}{(k-p+\epsilon)!}
=\lim_{\epsilon\rightarrow0}\sum_{k=p}^n\frac{\Gamma(k+1)\Gamma(\epsilon-p)}{\Gamma(k-p+\epsilon+1)\Gamma(\epsilon-p)}\nn\\%
&=&\lim_{\epsilon\rightarrow0}\frac{1}{\Gamma(\epsilon-p)}\sum_{k=p}^n\int_0^1
x^k(1-x)^{\epsilon-p-1}\nn\\%
&=&\lim_{\epsilon\rightarrow0}\frac{1}{\Gamma(\epsilon-p)}\int_0^1
(x^p-x^{n+1})(1-x)^{\epsilon-p-2}\nn\\%
&=&\lim_{\epsilon\rightarrow0}\frac{1}{\Gamma(\epsilon-p)}\big[\frac{\Gamma(p+1)\Gamma(\epsilon-p-1)}{\Gamma(\epsilon)}-
\frac{\Gamma(n+2)\Gamma(\epsilon-p-1)}{\Gamma(\epsilon-p+n+1)}\big]\nn\\%
&=&\frac{(n+1)!}{(n-p)!(p+1)}~.\nn\eea
This leads to%
\bea
&&\sum_{p=0}^{n-1}\sum_{k=p}^{n-1}\frac{k!(n-k+2)}{(n+3)!p!(k-p)!}
=\sum_{p=0}^{n-1}\frac{2p+n+5}{(p+2)!(n-p-1)!(n+1)(n+2)(n+3)}\nn\\%
&=&\frac{1}{2}\sum_{p=0}^{n-1}\frac{1}{(p+2)!(n-p+1)!}~,\nn\eea
where we take the average between $p\leftrightarrow n-p-1$ for the
last step. The final result is%
\bea \bar\partial_{[\bar j}R^n_{\bar i]l_1\cdots
l_{n+3}}=-\frac{1}{2}\sum_{p=0}^{n-1}\frac{1}{(p+2)!(n-p+1)!}R^p_{[\bar
j|l_1\cdots l_{p+2}m}\Omega^{mn}R^{n-p-1}_{\bar i]l_{p+3}\cdots
l_{n+3}n}+\textrm{perm in }l_i~.\nn\eea
Introducing formal variable $\xi^i$ which transforms as a vector we define
\bea \Theta_{\bar i} (x, \xi) \equiv \sum_{n=0}^{\infty}\frac{1}{(n+3)!}R^n_{\bar
il_1\cdots l_{n+3}}(x) \xi^{l_1}\cdots\xi^{l_{n+3}}~,\eea
which satisfies  the key identity%
\bea \bar\partial_{[\bar j}\Theta_{\bar i]}
=-\frac{1}{2}\{\Theta_{[\bar j},\Theta_{\bar i]}\}~,\label{flatw3293}\eea
 where $\{~,~\}$ stands for Poisson bracket in $\xi$-direction with respect to $\Omega_{ij}(x)$.
The equation (\ref{flatw3293}) is flatness condition and
hence there is an $L_{\infty}$ structure defined for a
hyperK\"ahler manifold.

Next we show that%
\bea \bar\partial_{\bar i}=\exp^{-1*}(\bar\partial_{\bar
i}+\{\Theta_{\bar i},\cdot\})\exp^*~.
\eea%
It should be understood that the right $\bar\partial$ acts on the
base of $T^{(1,0)\infty}_M$ and the left one acts on the second
factor of $M\times M$.

We need to show $(\partial_{\bar i}+\{\Theta_{\bar
i},\cdot\})\exp_{\xi}X^i=0$ on the diagonal (where $X^i=(X^{\bar
i})^*$), and this is equivalent to showing%
\bea e^{-\xi\cdot\nabla}(\partial_{\bar i}+\{\Theta_{\bar
i},\cdot\})e^{\xi\cdot\nabla}X^i=0;\hspace{.5cm}
\nabla_i:=\frac{\partial}{\partial X^i}-\xi^j\Gamma^k_{ij}\frac{\partial}{\partial\xi^k}\nn\eea%
By using the formula $e^{-A}Be^A=e^{-[A,\cdot]}B$, and the
flatness property $R_{ij\ \times}^{\ \ \times}=0$, we can show for example%
\bea &&e^{-\xi\cdot\nabla}\partial_{\bar
i}e^{\xi\cdot\nabla}=\partial_{\bar
i}+\sum_{n=1}^{\infty}\frac{(-1)^n}{n!}\xi^{i_1}\cdots\xi^{i_n}\Big(\xi^{i_{n+1}}\nabla_{i_1}\cdots
\nabla_{i_{n-1}}R_{\bar ii_{n}\ i_{n+1}}^{\;\ \
k}\partial_{\xi^k}\nn\\%
&&\hspace{4cm}-(n-1)\nabla_{i_1}\cdots \nabla_{i_{n-2}}R_{\bar
ii_{n-1}\ i_n}^{\;\ \ \ \ k}\nabla_k\big).\nn\eea%
Let us agree to write $\xi^{n}\nabla^{n-2}R_{\bar
i}^k:=\xi^{i_1}\cdots\xi^{i_n}\nabla_{i_1}\cdots
\nabla_{i_{n-2}}R_{\bar ii_{n-1}\ i_{n}}^{\;\ \ \ \ k}$, then%
\bea e^{-\xi\cdot\nabla}\big(\xi^{m+2}\nabla^mR_{\bar
i}^k\partial_{\xi^k}\big)e^{\xi\cdot\nabla}
=\sum_{n=0}^{\infty}\frac{(-1)^n}{n!}\Big(\xi^{n+m+2}\nabla^{n+m}R_{\bar
i}^{k}\partial_{\xi^k}-n\xi^{n+m+1}\nabla^{n+m-1}R_{\bar
i}^k\nabla_k\Big),\nn\eea%
And combining the two,%
\bea e^{-\xi\cdot\nabla}\big(\bar\partial_{\bar i}+\{\Theta_{\bar
i},\cdot\}\big)e^{\xi\cdot\nabla}&=&e^{-\xi\cdot\nabla}\big(\bar\partial_{\bar
i}+\sum_{m=0}^{\infty}\frac{1}{(m+2)!}\xi^{m+2}\nabla^mR_{\bar
i}^k\partial_{\xi^k}\big)e^{\xi\cdot\nabla}\nn\\%
&=&\bar\partial_{\bar
i}+\sum_{n=1}^{\infty}\frac{(-1)^n}{(n+1)!}\xi^{n+1}\nabla^{n-1}R_{\bar
i}^k\partial_{\xi^k},\nn\eea%
This shows clearly that $(\partial_{\bar i}+\{\Theta_{\bar
i},\cdot\})\exp_{\xi}X^i=0$.

\end{document}